\documentclass[sigconf, nonacm]{acmart}

\AtBeginDocument{%
  \providecommand\BibTeX{{%
    Bib\TeX}}}

\setcopyright{acmcopyright}
\copyrightyear{2023}
\acmYear{2023}
\acmDOI{XXXXXXX.XXXXXXX}

\AtBeginDocument{%
  \providecommand\BibTeX{{%
    \normalfont B\kern-0.5em{\scshape i\kern-0.25em b}\kern-0.8em\TeX}}}

\usepackage{natbib} 
\usepackage{mathtools} 
\usepackage{booktabs} 
\usepackage{tikz} 
\usepackage[mode=buildnew]{standalone}

\usepackage{caption}
\usepackage{subcaption}

\usepackage{todonotes, xcolor}

\begin{document}



\title{Human Uncertainty in Concept-Based AI Systems}


\author{Katherine M. Collins}
\affiliation{%
  \institution{University of Cambridge}
  \country{United Kingdom}
}
\email{kmc61@cam.ac.uk}

\author{Matthew Barker}
\authornote{Contributed equally (sorted alphabetically by last name).}
\affiliation{%
  \institution{University of Cambridge}
  \country{United Kingdom}
}

\author{Mateo Espinosa Zarlenga}
\authornotemark[1]
\affiliation{%
  \institution{University of Cambridge}
  \country{United Kingdom}
}

\author{Naveen Raman}
\authornotemark[1]
\affiliation{%
  \institution{University of Cambridge}
  \country{United Kingdom}
}

\author{Umang Bhatt}
\affiliation{%
  \institution{University of Cambridge \\ Alan Turing Institute}
  \country{United Kingdom}
}

\author{Mateja Jamnik}
\affiliation{%
  \institution{University of Cambridge}
  \country{United Kingdom}
}

\author{Ilia Sucholutsky}
\affiliation{%
  \institution{Princeton University}
  \country{United States}
}

\author{Adrian Weller}
\affiliation{%
  \institution{University of Cambridge \\ Alan Turing Institute}
  \country{United Kingdom}
}

\author{Krishnamurthy (Dj) Dvijotham}
\affiliation{%
  \institution{Google Research (Brain Team)}
  \country{United States}
}

\renewcommand{\shortauthors}{Collins, et al.}


\begin{abstract}
Placing a human in the loop may abate the risks of deploying AI systems in safety-critical settings ( e.g., a clinician working with a medical AI system). However, mitigating risks arising from human error and uncertainty within such human-AI interactions is an important and understudied issue. 
In this work, we study human uncertainty in the context of concept-based models, a family of AI systems that enable human feedback via concept interventions where an expert intervenes on human-interpretable concepts relevant to the task. 
Prior work in this space often assumes that humans are oracles who are always certain and correct. Yet, real-world decision-making by humans is prone to occasional mistakes and uncertainty. We study how existing concept-based models deal with uncertain interventions from humans using two novel datasets: \texttt{UMNIST}, a visual dataset with controlled simulated uncertainty based on the MNIST dataset, and \texttt{CUB-S}, a relabeling of the popular \texttt{CUB} concept dataset with rich, densely-annotated soft labels from humans. We show that training with uncertain concept labels may help mitigate weaknesses of concept-based systems when handling uncertain interventions. These results allow us to identify several open challenges, which we argue can be tackled through future multidisciplinary research on building interactive uncertainty-aware systems. To facilitate further research, we release a new elicitation platform, \texttt{UElic}, to collect uncertain feedback from humans in collaborative prediction tasks. 
\end{abstract}





 \maketitle


\section{Introduction}

\begin{figure*}[!htb]
    \centering
    \includegraphics[width=1.0\linewidth]{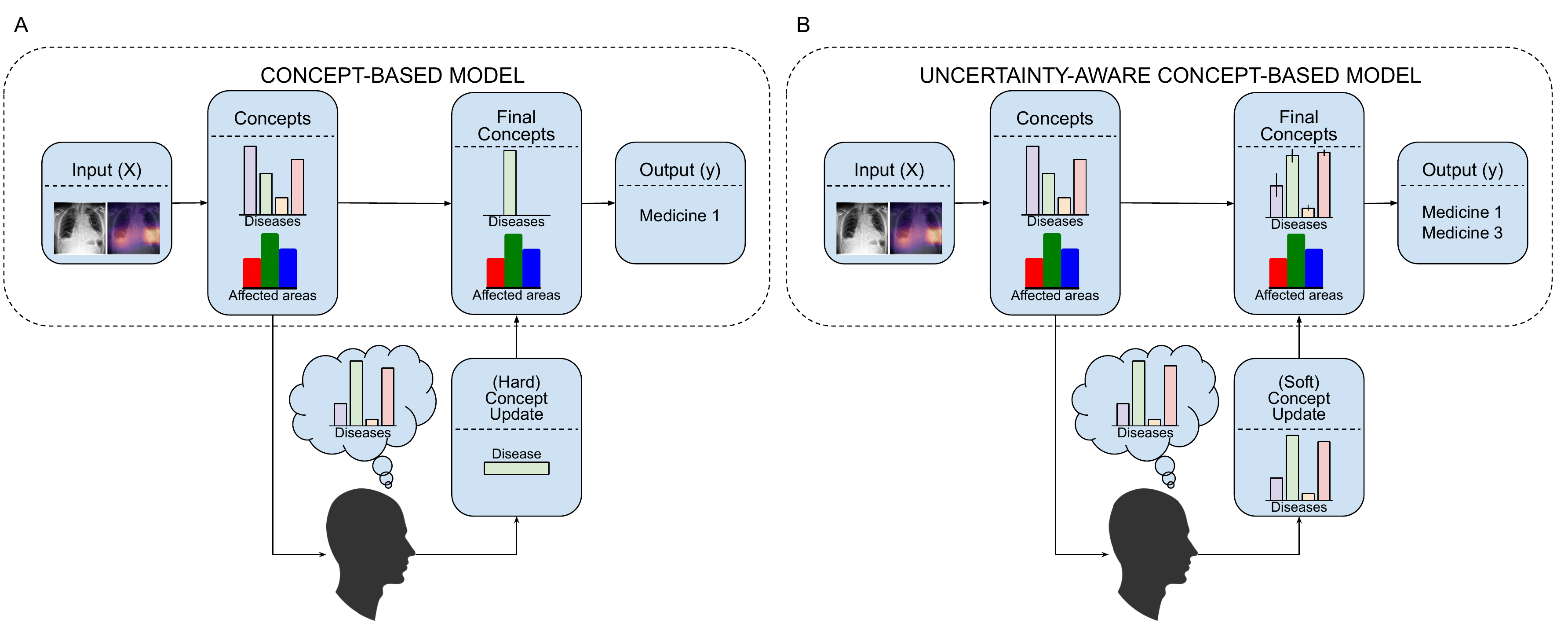}
    \caption{Schematic of uncertainty in test-time interventions in concept-based models. When presented with features and a concept to annotate, a human user may be uncertain. We empower the user to express this uncertainty when intervening on concepts: to make the concept-based systems \textit{aware of their uncertainty}. We demonstrate the set-up in a hypothesized safety-critical setting of medical diagnosis; X-ray images are depicted from CheXpert \cite{irvin2019chexpert}.}
    \label{fig:main-schematic}
\end{figure*}

Human-in-the-loop machine learning (ML) systems are often framed as a promising way to reduce risks in settings where automated models cannot be solely relied upon to make decisions \citep{shneiderman2022human}. However, what if the humans themselves are unsure? Can such systems robustly rely on human interventions which may be inaccurate or uncertain? Concept-based models (e.g., Concept Bottleneck Models (CBMs) \citep{koh2020concept} and Concept Embedding Models (CEMs) \citep{cem22}), are ML models that enable users to improve their predictions via feedback in the form of human-interpretable ``concepts'', as opposed to feedback in the original feature space (e.g., pixels of an image). For instance, a radiologist can identify concepts like lung lesions or a fracture to aid a model which uses chest X-rays to predict diseases. Such human-in-the-loop systems typically assume that the intervening human is always correct and confident about their interventions; a so-called ``oracle'' whose predictions should always override those of the model (see Figure~\ref{fig:main-schematic}A). Yet, uncertainty is an integral component of the way humans reason about the world ~\citep{uncertainJudgments, Gha15, probBrain, lake_ullman_tenenbaum_gershman_2017}. If a doctor is unsure of whether a lung lesion is present, or a human cannot observe a feature in a bird due to occlusion (e.g. the tail of a bird is hidden from view), it may be safer to permit them to express this uncertainty \citep{medUncertainty, laidlaw2021uncertain, schneider2022effects}. Human-in-the-loop systems, which can take uncertainty into account when responding to human interventions, would mitigate the risks of both end-to-end automation and human error (see Figure~\ref{fig:main-schematic}B).



Just as machines ``knowing when they don't know'' has been emphasized for reliability \citep{bhatt2021uncertainty, liu2020simple, uncTrustModel, hendrycks2019benchmarking}, we emphasize \texttt{empowering humans to express when they do not know} as a way to improve trustworthy deployment and outcomes. Recent works have demonstrated the benefits of incorporating uncertainty over label spaces on predictive performance \citep{hinton2015distilling, muller2019labelSmoothHelp, peterson2019human, selfCiteSoftLabel,informativenessSoft,hillmixup,sandersambiguous}; we continue this tradition in the space of concept-based \textit{feedback}. Specifically, our contributions can be summarized as follows:

\begin{itemize}
    \item We introduce the safety-critical problem of human uncertainty in interactive, concept-based models. 
    \item We reveal failure modes of existing concept-based models when handling user uncertainty over concepts.
    \item We empirically demonstrate the value of training with uncertainty as a mitigation strategy for better handling test-time uncertainty. 
    \item We develop \texttt{UElic}, an extensible platform to facilitate the collection of rich, real-world human uncertainty over concepts. 
    \item We use \texttt{UElic} to curate a novel relabeling of \texttt{CUB} (called \texttt{CUB-S}) designed to address limitations in the present dataset. Furthermore, we illustrate how \texttt{CUB-S} can serve as a challenge dataset to study uncertain human interventions. 
\end{itemize}

\section{Primer on Concept-Based Systems}

In this section, we introduce concept-based models and discuss how their design enables concept interventions. Concept-based models use human-interpretable values (concepts) as intermediate representations when predicting a task label~\citep{koh2020concept}. The aim of such models is to improve the interpretability of the outputs and to facilitate human interventions which correct model mistakes~\citep{koh2020concept,chen2020concept, barbiero2022entropy, cem22, chauhan2022interactive}.

\subsection{Notation}
We consider the supervised case where each datapoint consists of (input $\mathbf{x} \in \mathcal{X}$, concepts $\mathbf{c} \in \mathcal{C}$, output $\mathbf{y} \in \mathcal{Y}$). Typically, concepts $\mathbf{c} = [c_1, c_2, \cdots, c_k]^T$ are binary (indicating that concept is ``on'' or ``off''; e.g., oedema is or is not present), or categorical (e.g., different wing colors). Notice, however, that categorical concepts can be converted into binary concepts (e.g., wing color is or is not blue). Typically, concept presence is annotated as being ``on'' or ``off'' ($c_i \in \{0, 1\}$); however, there may be \textit{uncertainty} over a concept's presence, which necessitates a continuous value. For that reason, in this work, we let concepts live $\in [0,1]$, representing $p(c_i | x)$.  

\subsection{Models}
Concept-based models predict the concepts from an intermediate layer in a neural network. Although a plethora of such systems have been developed \citep{koh2020concept, cem22, postHocCBMs, conceptSidecar, languageCBM, oikarinenlabel}, in this work we focus on Concept Embedding Models (CEMs) \citep{cem22} as they represent a recent extension of the popular Concept Bottleneck Models (CBMs) \citep{koh2020concept}. 

CBMs learn two mappings, one from the input to the concepts $g: \mathcal{X} \rightarrow \mathcal{C}$, and another from the concepts to the outputs $f: \mathcal{C} \rightarrow \mathcal{Y}$. The overall prediction is given by:
\begin{equation}
    \hat{y} = f(\mathbf{\hat{c}}) = f(g(\mathbf{x}))
\end{equation}
There are many ways of learning $g$ and $f$; here, we focus on the joint bottleneck, which learns $g$ and $f$ at the same time, simultaneously minimizing the concept prediction loss and the output prediction loss. In this work we focus on CBMs with sigmoidal activations in their concept layers whose output can be interpreted as a concept's probability of activation. CEMs further extend CBMs by learning supervised embeddings for each concept, representing concepts as high dimensional vectors while still learning to predict their values as an intermediate step~\citep{cem22}. This allows CEMs to better leverage their capacity when trained on datasets with an ``incomplete'' set of concept annotations~\citep{cem22, yeh2020completeness}. We use \textbf{training with uncertainty} to refer to models trained with concepts represented as probabilities $\in [0,1]$, rather than as binary concepts. The target $y$ is left unchanged in this work. 




\subsection{Interventions}

A prime motivation for employing concept-based systems is the ease of intervenability. If a user notices that the model is predicting a concept incorrectly (e.g., the X-ray scan shows bone spurs, yet the model predicted no bone spurs), a user (e.g., a medical professional) can directly edit said concept to (potentially) update the prediction. This involves updating a predicted concept $\hat{c}_i$ with the concept value returned by the human $\hat{c}_i \leftarrow c_i$ and recalculating our prediction $\hat{y} = f(\mathbf{\hat{c}})$. Because these interventions edit the \textit{model's predicted probability of a given concept}, we can readily permit the user to edit \textit{with their own predicted probability of that concept}. When we refer to \textbf{testing with uncertainty}, we let the human edited concept be a probability, analogous to our ``training with uncertainty'' setting.

In this work, we consider two policies to select the concept to intervene on: 1) \textit{Random}: selecting the next concept to query randomly of concepts, and 2) \textit{Skyline:} an approximate oracle policy following \citeauthor{chauhan2022interactive}, which selects the next concept to query that will best impact performance (as if it were possible to know, simulating an upper bound on intervention efficacy; see Supplement for further details). While other works have been developed with more advanced policies \citep{chauhan2022interactive, sheth2022learning, shin2023closer}, we select Random and Skyline because they illustrate the \textit{bounds} on achievable performance; Random being the most naive policy and Skyline being the optimal policy. Unless otherwise noted, concepts are chosen via Random. 

\subsection{Critiques and Common Assumptions} Concept-based models, and the broader ecosystem in which they are deployed, have been shown to exhibit information leakage~\citep{mahinpei2021promises} or impurities distributed across concept representations~\citep{zarlenga2023towards}, spurious input saliency maps \citep{margeloiu2021concept}, bloated, hard-to-learn concept definitions \citep{ramaswamy2022overlooked}, and propensity to be influenced by correlations amongst concepts \citep{heidemann2023concept}. To our knowledge, we are the first work which directly considers \textit{uncertainty in the human user} with concept-based models.












\section{Research Questions}

In this work, we address the following research questions: 

\begin{itemize}
    \item \textbf{RQ1}: How do existing concept-based systems handle the introduction of human uncertainty at test time?
    \item \textbf{RQ2}: How can systems be bolstered to better support human uncertainty at test time? 
    \item \textbf{RQ3}: How does the level and form of the uncertainty (e.g., whether the uncertainty is expressed through discrete annotations, or rich, continuous probabilities) impact performance? 
\end{itemize}

We investigate these questions across a \textit{spectrum of degrees and forms of uncertainty}. First, we study controlled, simulated uncertainty in \texttt{UMNIST}, our newly proposed addition dataset based on the \texttt{MNIST} handwritten digit recognition, as well as over the popular medical dataset \texttt{CheXpert} \citep{irvin2019chexpert}. We then depart from considering simulated uncertainty -- moving moving to the real, human-elicited in-the-wild uncertainty; first, coarse-grained uncertainty scores collected in \texttt{CUB} \citep{WahCUB_200_2011}, and then richer uncertainty which we collect in our new real-world dataset of human uncertainty: \texttt{CUB-S}.  

For each dataset, we study the test-time performance of models trained on binary, certain concepts, but faced with uncertainty at test-time. Then, we explore how this performance is affected when the same models are trained with uncertainty estimations in concept labels.

\section{Simulated Uncertainty}

We first investigate the performance of concept-based models on simulated uncertainty.


\subsection{Experimental Set-Up}

\subsubsection{Data} We consider two datasets with varying degrees of simulated uncertainty: \texttt{CheXpert} and a newly constructed, controllable dataset of uncertainty, \texttt{UMNIST}. \texttt{CheXpert} is a visual dataset containing chest X-rays that are annotated with a set of 14 concepts. In our scenario, we aim to predict the ``No Finding'' concept based on the other 13 concepts. We incorporate simulated uncertainty by each concept's label by setting uncertain values to 0.5 and unknown values to 0 (\texttt{CheXpert} comes with annotations indicating which concepts are uncertain/unknown). \texttt{UMNIST}'s samples are formed by a mixture of MNIST digits (e.g., zeros or ones), where the task is to compute the sum of all digits and each sample is given the number represented by each digit as a concept annotation (see Supplement for more details). \texttt{UMNIST} is parameterized with parameter $\delta \in [0, 1]$ which controls the amount of uncertainty/noise in each sample's concept annotations. Intuitively, $\delta =0$ represents fully certain concept labels and no mixing of each sample's digits while $\delta = 1$ represents a dataset with random concept annotations. Uncertainty in the \texttt{UMNIST} dataset, therefore, is applied uniformly over concepts, at varying levels. We apply the same $\delta$-smoothing to \texttt{CheXpert}.


\subsubsection{Evaluation} 

We study the performance of the concept-based systems on the task of interest (e.g., abnormality detection in chest X-rays, predicting the sum of digits in an image) as a function of the number of concepts intervened. For both \texttt{CheXpert}, following \citeauthor{chauhan2022interactive}, we evaluate the Area under the ROC curve (\textit{AUC}). For \texttt{UMNIST}, given its multi-class setting, we evaluate accuracy instead. Finally, as we are interested in how uncertain interventions affect concept-based models rather than how to best take into account uncertainty at intervention time, an interesting yet different research question, in our evaluation we randomly choose which concepts to intervene on rather than deploying more principled intervention policies.

\subsection{Intervening with Uncertainty}

\begin{figure}[!htb]
    \centering
    \includegraphics[width=1.0\linewidth]{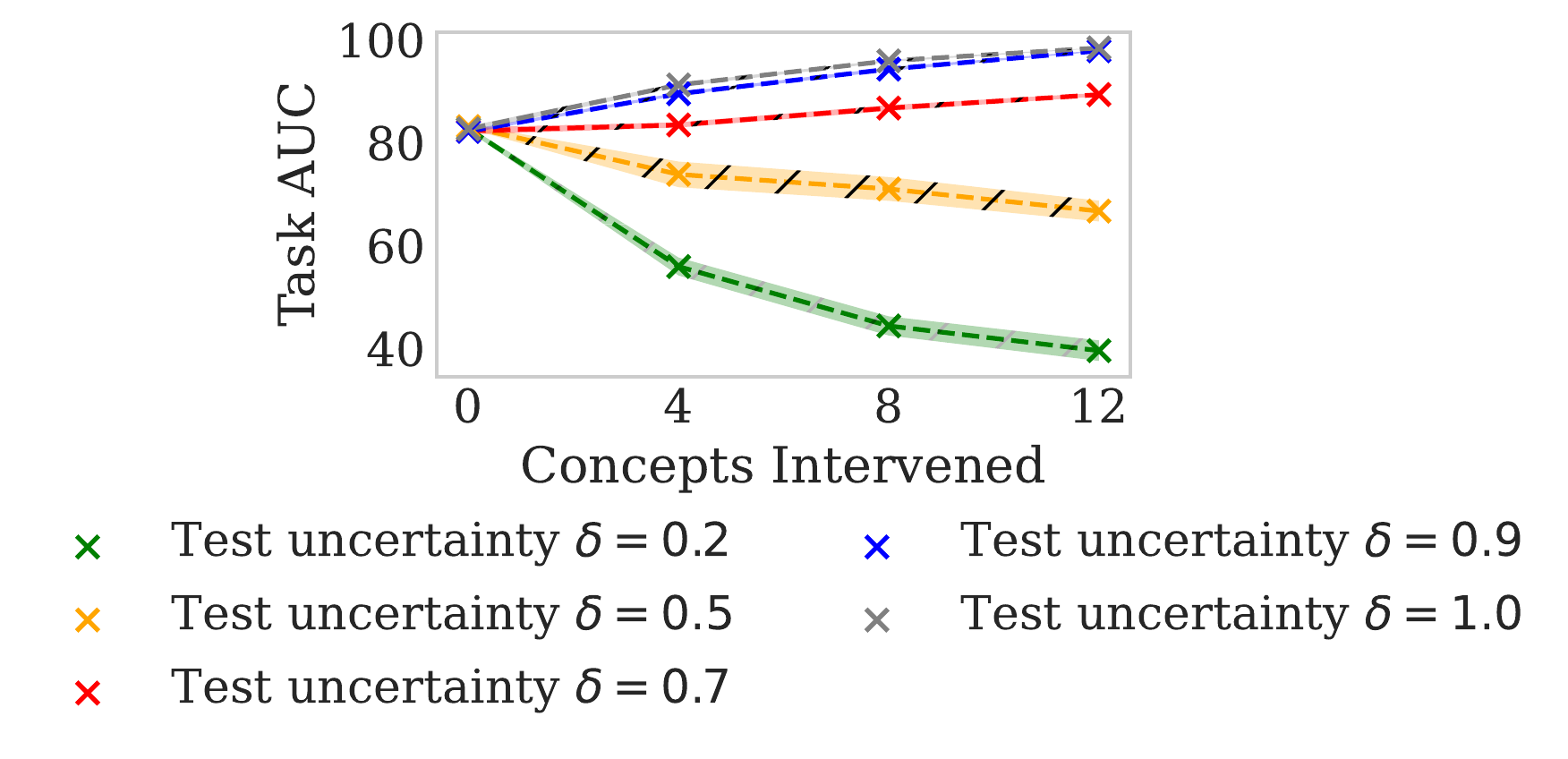}
        \caption{Effect of simulated uncertainty in \texttt{CheXpert} on test-time efficacy (AUC) in CME as the number of concepts intervened on increases. Standard error depicted over 3 random seeds.}
    \label{fig:chexpert_sim_intervention}
\end{figure}

\begin{figure*}[!htb]
    \centering
    \includegraphics[width=0.47\linewidth]{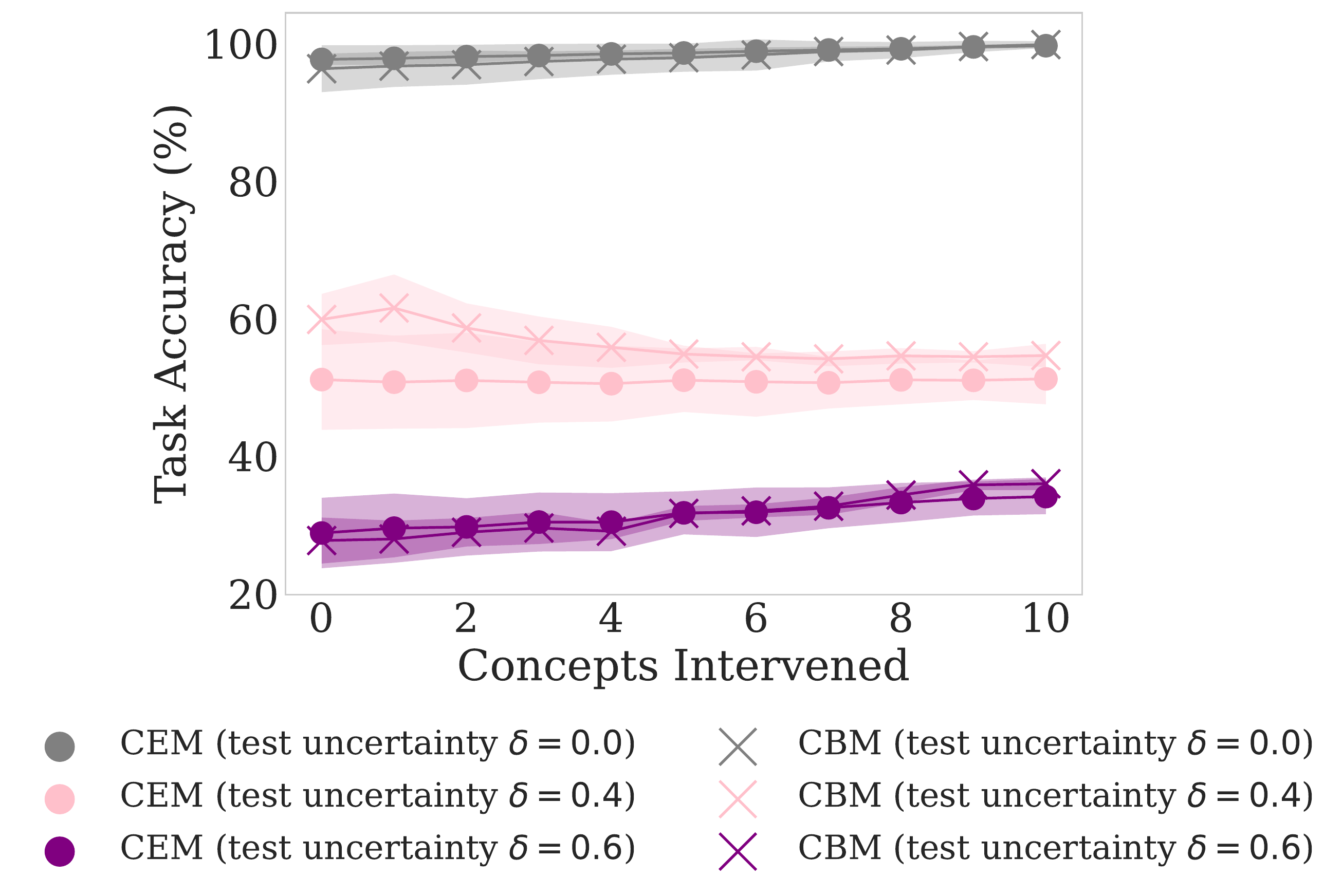}
    \includegraphics[width=0.43\linewidth]{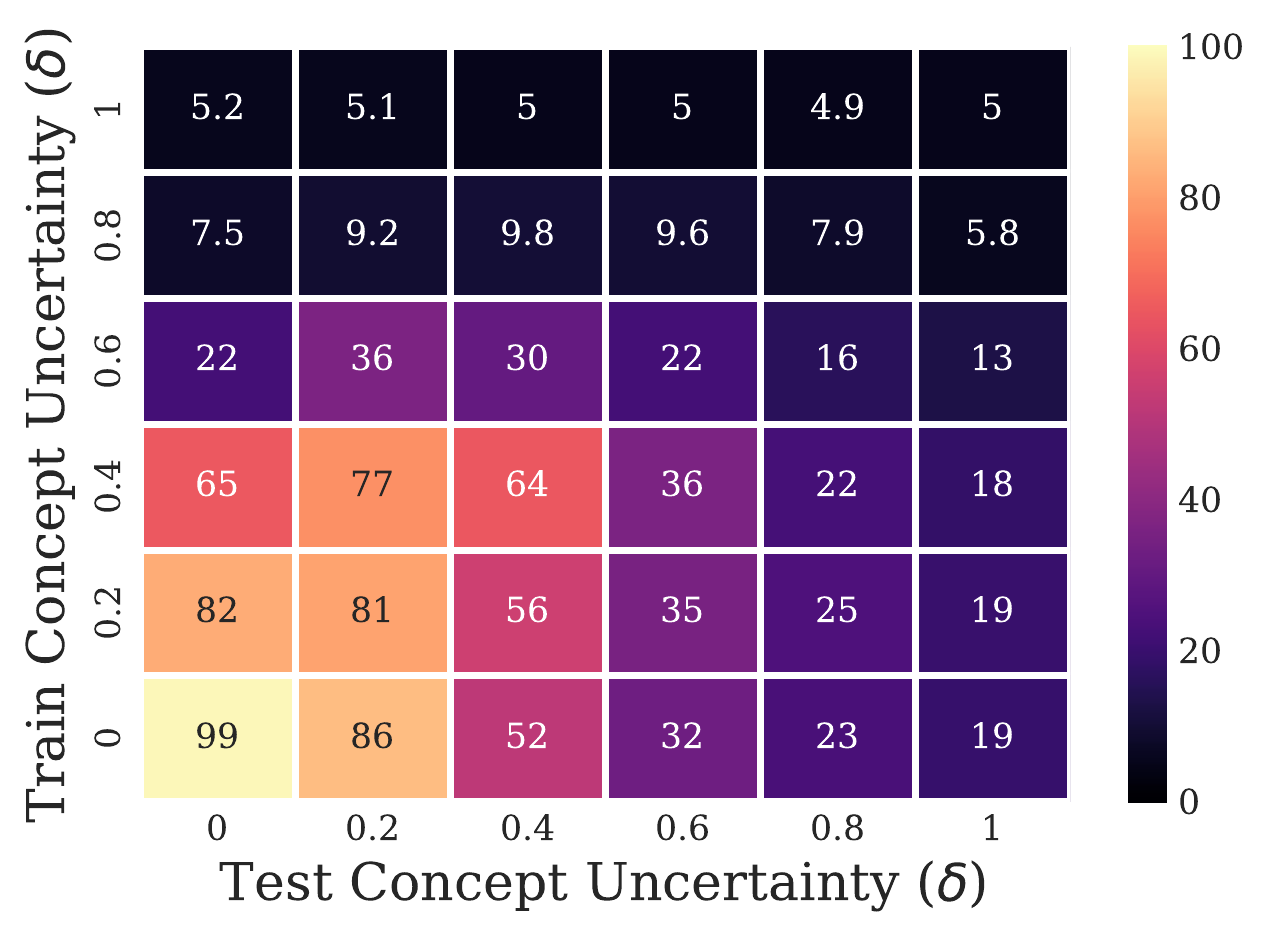}
    \caption{\textbf{Left}: Mean test accuracies of random interventions on CBMs and CEMs, together with their standard error computed across 5 different random initializations, as we increase the number of concepts we intervene on. Concept-based systems (CBMs and CEMs) that have not been trained on uncertainty struggle to handle test-time uncertainty, even when both models achieved similarly high concept accuracies.  \textbf{Right}: Heatmap showing the task accuracy (\%) of a CEM trained in \texttt{UMNIST} (with train-time $\delta$ varying across the y-axis) after intervening in 50\% of its concepts with possibly uncertain test-time concept labels (controlled by the test dataset's $\delta$ value in the x-axis). Training with uncertainty in \texttt{UMNIST} improves robustness under distribution shift at intervention time (compare bottom row when test time $\delta \in \{0.4, 0.6\}$ vs CEMs trained with samples generated with $\delta \in \{ 0.2, 0.4 \}$), provided the training level of uncertainty is not too high. 
    }
    \label{fig:umnist_res}
\end{figure*}

We first benchmark how well models \textit{not trained with uncertain concepts} cope with uncertainty at intervention during testing. This setting best captures what a user facing uncertainty may experience when deploying pre-trained concept-based models, which are rarely trained on uncertainty. Specifically, we explore varying the total amount of concept uncertainty in the testing data and observe that, even with low simulated uncertainty, both CEMs and CBMs suffer from significant drops in intervention performance when dealing with uncertain samples (see Figures~\ref{fig:chexpert_sim_intervention}~and~ \ref{fig:umnist_res}; see Supplement). We can see that this drop is particularly sharp as the amount of uncertainty grows, as seen in the performance of CEMs when $\delta = 0.6$. This suggests that these models, although accurate and high performing when receiving fully certain interventions, are unable to generalize to settings in which the intervening user is uncertain of the nature of some of the concepts.

\subsection{Training with Uncertainty Can Improve Robustness} 

While we observe that exposing models not trained with concept uncertainty to uncertain concepts leads to the breakdown of intervention efficacy --- we hypothesize that training with uncertainty can boost the ability to cope. We indeed observe in Figures \ref{fig:umnist_res} (Right) and \ref{fig:chexpert_sim_intervention} that by \textit{training} with uncertainty, we can salvage the efficacy of interventions -- particularly under \textit{distribution shift}. That is, if we train on an uncertainty level that differs from the level expressed by a user, we may be better equipped to handle that user's uncertainty than if we did not train with uncertainty at all. Notably, however, we observe a ``sweet spot'' in the level of uncertainty that is helpful to the model. 


\subsection{Implications} Even in controlled settings, existing concept-based systems struggle to adequately handle concept uncertainty at inference time. Training with concept uncertainty proves a reasonable salve for capturing value from the uncertain interventions, particularly affording robustness under distribution shifts. However, our results suggest that training with too much concept ``softness'' can be harmful.

\section{Real human uncertainty}



We see in our simulations that permitting test-time uncertainty can impact model performance --- and that training with uncertainty offers a potential remediation strategy when handling test-time uncertainty. However, these investigations are on contrived uncertainty, begging to ask how existing systems fare with real-world uncertainty.

\subsection{Taxonomy of Forms of Uncertainty}

Real human uncertainty can come in many forms. This uncertainty may be \textbf{epistemic}, representing lack of knowledge, or \textbf{aleotoric}, due to (potentially) inevitable randomness \citep{hullermeier2021aleatoric}. Further, this uncertainty can either be \textbf{heteroschedastic}, i.e., dependent on the input, or \textbf{homeoschedastic}, independent of the input \citep{books/lib/RasmussenW06}. Thus far, we have focused on \textit{regular} uncertainty -- simulating the same level of uncertainty $\delta$ across all concepts. 

However, in-the-wild uncertainty, elicited from humans, is not so simple. The method by which uncertainty is elicited can have a sizeable impact on the quality of the elicitation \citep{uncertainJudgments, goldstein2014lay, keren1991calibration, expertElicitation}. As researchers may use a variety of elicitation practices, we believe it is \textbf{important to understand how concept-based systems handle different forms of elicited human uncertainty}. 

We focus on two flavors of uncertainty \textbf{coarse-grained} (elicited from a few-option discrete scale) and \textbf{fine-grained} (probabilities extracted over each possible attribute in a concept group). In the coarse-grained setting, humans provide both binary concept annotations, $c_i \in \{0, 1\}$, and a discrete measure of confidence $\omega$, e.g., $\omega \in \{$``Guessing'', ``Probably'', ``Definitely''\}. In this setup, we need to construct a map from $c_i \times \omega$ to the probability distribution of interest $p(c_i | x)$. In contrast, in the fine-grained setting, humans directly provide $p(c_i | x)$. 

While we do not consider \textit{all} forms of uncertainty expression, e.g., humans may prefer to express uncertainty flexibly through language \citep{DHAMI2022514, zhou2023navigating}, we see our study as a promising first step into a deeper investigation of the impact of different forms of \textit{real human uncertainty} on concept-based system performance. 


\subsection{Coarse-Grained Uncertainty}

We first consider these questions over \textit{coarse-grained} human uncertainty; i.e., a single discrete annotation of whether the user is uncertain. The limited, discrete nature of the uncertainty variable $\omega$ raises important design considerations when considering how to use this uncertainty. For instance, if a user marks that they are uncertain, how can we know \textit{how} uncertain are they? And are they only uncertain over parts or the entirety of the concept space? We next study how design choices to impute these ambiguities at train- and test-time can impact the intervention efficacy. We address these questions through the uncertainty annotations provided in \citep{WahCUB_200_2011}.\footnote{We include analyses over the ``real''  coarse-grained uncertainty annotations in \texttt{CheXpert} in the Supplement. In contrast to \texttt{CUB}, which has uncertainty annotations for each image and attribute, only some concepts are labeled with an uncertainty score in CheXpert. Moreover, the score obfuscates whether the label is deemed uncertain due to human uncertainty versus annotation-scraping uncertainty \citep{irvin2019chexpert}. As such, we place less emphasis on CUB in this work.}.

\subsubsection{Experimental Set-Up}

\paragraph{Data}

CUB is a highly popular benchmark dataset for concept prediction that includes images of birds, annotated with 28 different concept groups (e.g., wing color, beak shape) \citep{WahCUB_200_2011}. Each concept can take on many different values. The task is to predict one of two hundred different bird species. \citeauthor{WahCUB_200_2011} elicited humans' uncertainty when collecting the original annotations; however, these annotations are highly coarse (a simple: ``Guessing,'' ``Probably'', or ``Definitely'' mark over each concept group's annotations). There are 311 total binary concepts that can be extracted from the 28 categorical concepts; we follow the common practice proposed in \citeauthor{koh2020concept} by filtering these down to 112 concepts. We study how intervening with, and learning with, these coarse-grained annotations impacts performance. 


\paragraph{Evaluation}

We follow similar evaluation protocols to our Simulated Uncertainty experiments, focusing on the measure of task accuracy (where the task is bird species classification). We include Skyline interventions to help demonstrate the best possible intervention policy that can be achieved to further highlight the impact of the types of uncertainty on performance.

\subsubsection{How to Use Discrete Uncertainty Scores?} 

The first question raised with the real-world uncertainty of the form elicited in \texttt{CUB} is how to leverage the scores at intervention time. Uncertain annotations are only provided in the form of a single, discrete measure of uncertainty: \texttt{CUB} annotators provided \textit{coarse-grained}, discretized approximations of their confidence in said annotations (i.e., specifying whether they were Guessing, Probably Sure, or Definitely Sure in their annotations). 

However, concept-based systems typically necessitate interventions to be specified in continuous space; as such, we need to define a custom mapping from discretely expressed uncertainty to continuous values. The choice of such a mapping impacts downstream performance. Second, for categorical concepts like those in \texttt{CUB}, a single measure of uncertainty does not permit a nuanced assignment of uncertainty to individual concepts. There is ambiguity around what the human user intended to express; i.e., if the user says they are ``Probably'' we do not know over which concepts and \textit{how} unsure they are.

We highlight the ramification of this ambiguity in two ways. First, we demonstrate that imputing the coarse-grained uncertainty with different continuous values can -- at times dramatically -- impact performance. Second, we demonstrate that the degree of softness assumed when leveraging uncertainty over \textit{categorical} concept spaces matters.   

\paragraph{Imputing the ``Probably'' Probability}

\begin{figure}[!htb]
    \centering
    \includegraphics[width=0.9\linewidth]{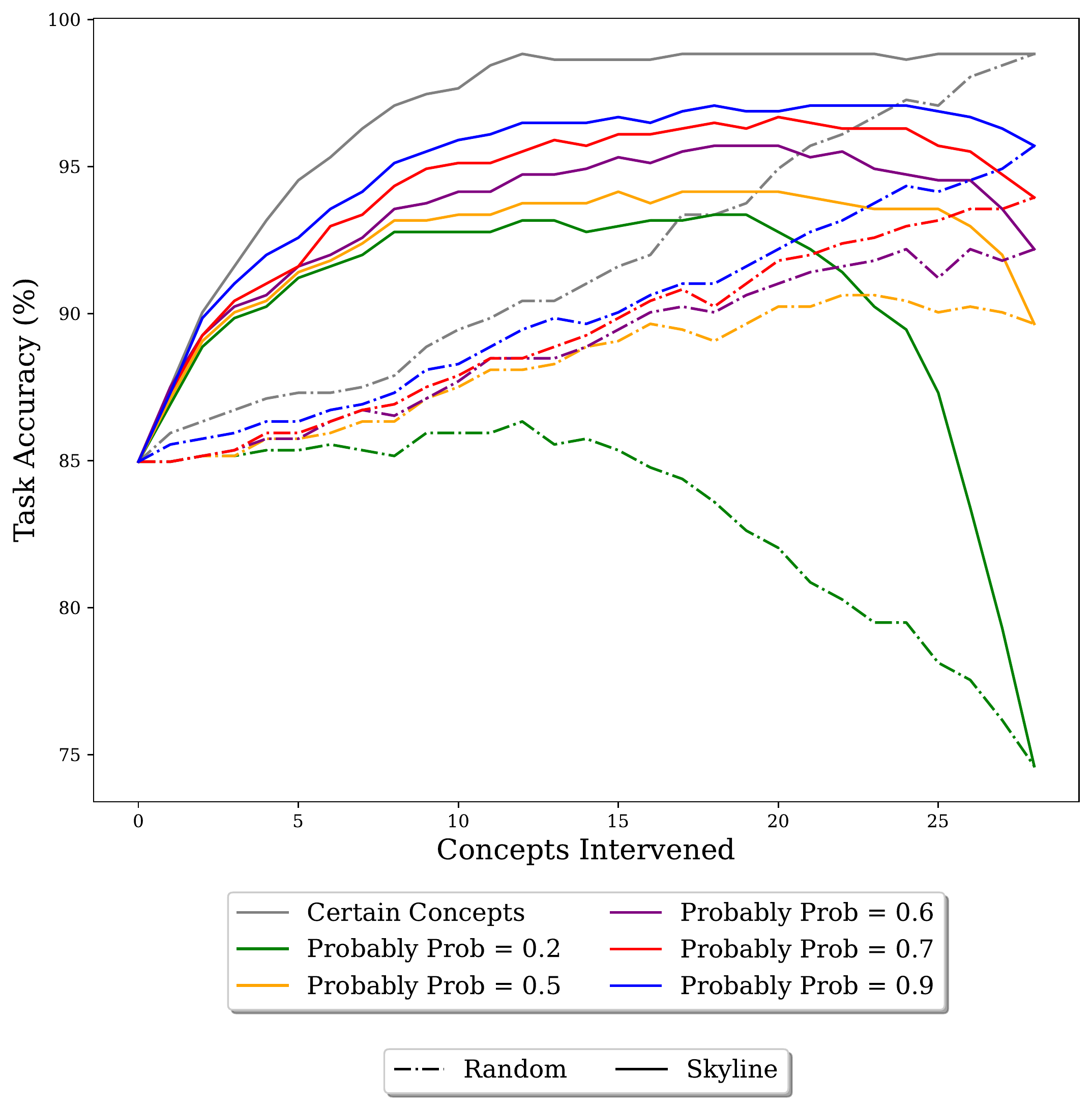}
    \caption{ Impact of different levels of uncertainty on intervention efficacy (task accuracy) in CEMs as the number of concepts intervened on increases, across both Random and Skyline policies. Colors correspond to the different imputations of the probability someone intends when they say they are ``Probably'' sure that are used at intervention-time. Mean performance when intervening over all test set examples in \texttt{CUB}.}
    \label{fig:vary_prob}
\end{figure}

We focus on the concept annotations where humans expressed they were ``Probably'' sure of the annotations. Here, we do not know \textit{how} certain the annotators were in their labeling. We vary the level of uncertainty we assume annotators were in such annotations when intervening and apply the same imputed probably to the ``on'' (e.g., blue wing present) and ``off'' concepts (e.g., wing color is not yellow); for the latter, we flip the assigned probability. We observe in Figure \ref{fig:vary_prob} that the imputed probability can have a dramatic impact. The imputation matters -- demonstrating both limitations of insufficient richness in annotation (we do not know what the original annotators intended) and further brittleness of these systems to test-time uncertainty when they have been trained exclusively on deterministic concepts.

\paragraph{Distribution of Uncertainty over Categorical Concepts} 

\begin{figure}[!htb]
    \centering
    \includegraphics[width=0.9\linewidth]{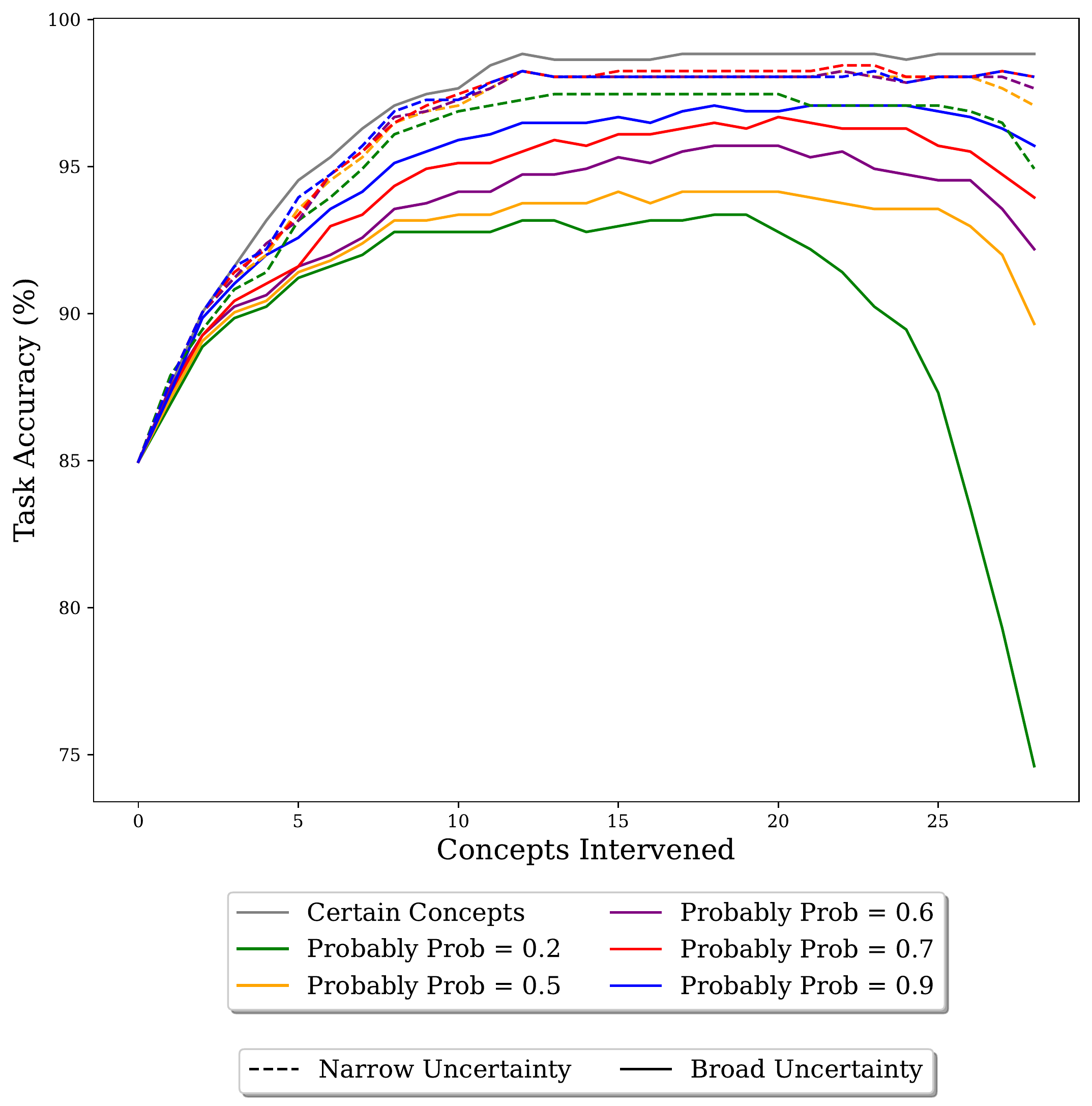}
    \caption{Impact on task accuracy of different ways of distributing the discrete uncertainty over categorical concept groups selected using Skyline.}
    
    \label{fig:skyline_smooth_off}
\end{figure}

Not only does insufficient richness in uncertain annotations pose a challenge when determining what level of certainty to assign: it is also ambiguous \textit{which} concepts the annotator was uncertain in when they said they were ``probably'' sure. We refer to this phenomenon as whether the annotator's uncertainty is \textbf{broad} (over all possible concept values) or \textbf{narrow} (just over a few of the possible concept values). For instance, when annotating beak shape, the annotator may be very certain the shape is not rounded -- but unsure whether to classify the shape as dagger-like or pointed: ``narrow'' uncertainty. In that case, the intervention on rounded should be left fully ``off'' (i.e., 0\%), but the mass should be spread on the possible ``on'' values (perhaps 70\% dagger, 30\% pointed). We demonstrate in Figures \ref{fig:skyline_smooth_off} and in the Supplement that these choices also matter. Assuming that an annotator's uncertainty is broad, and only over aspects of the concept space, can substantially impair intervention quality (likely because the converse was oversmoothing -- i.e., falsely miscalibrating to be underconfident). The sensitivity of the models and policies to these varied degrees of uncertainty highlights the brittleness of systems to such design choices and possible spectra of human uncertainty expression. 

\subsubsection{Instance- vs. Population-Level Uncertainty?}

Another question raised by in-the-wild human uncertainty is how to handle individual differences versus group-level uncertainty \citep{peterson2019human, selfCiteSoftLabel}. This question is particularly pertinent in \texttt{CUB}, as the annotations are both sparse and noisy. Several concepts have few annotations, and many annotations may be low-quality. As such, it may make sense to consider \textit{population-level} uncertainty rather than individual uncertainty. Here, we refer to population-level uncertainty as the class-level labels used by \citeauthor{koh2020concept}. We form soft labels by aggregating all annotators' individual-level soft labels for a given category. To ``upper bound'' the differences in population vs. individual-level uncertainty, we consider the possibility that annotators are unsure over \textit{both} ``on'' and ``off'' annotations (i.e., that they possess broad uncertainty). We see that whether or not to intervene with population-level uncertainty matters --- test-time performance is markedly higher when using population-level labels (see Supplement). 

\subsubsection{Training with Uncertainty}

\begin{figure*}[!htb]
\minipage{0.32\textwidth}
  \includegraphics[width=\linewidth]{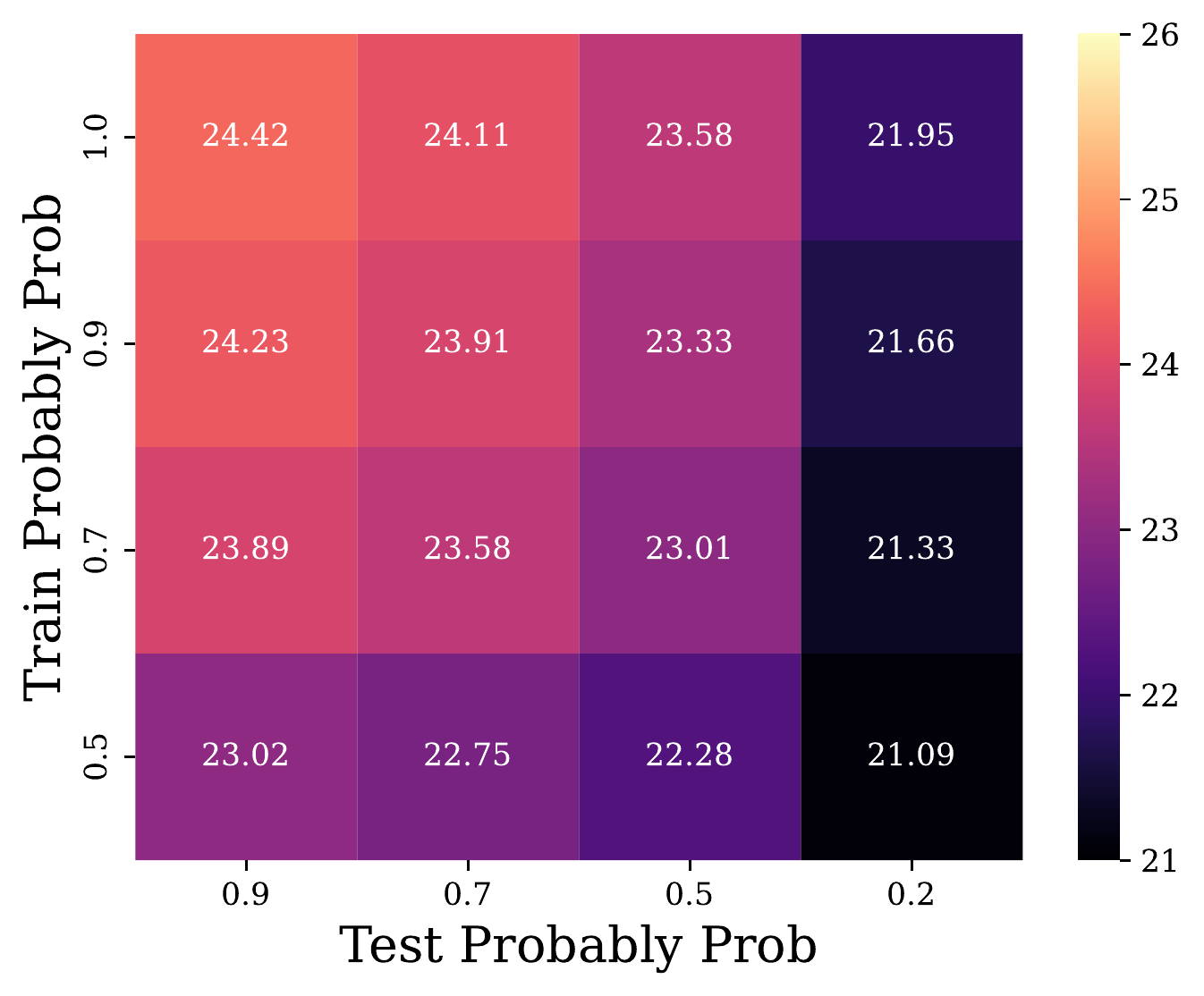}
  \subcaption{Training on instance-level (broad) uncertain concept annotations.}\label{fig:smooth_train_inst_test}
\endminipage\hfill
\minipage{0.32\textwidth}
  \includegraphics[width=\linewidth]{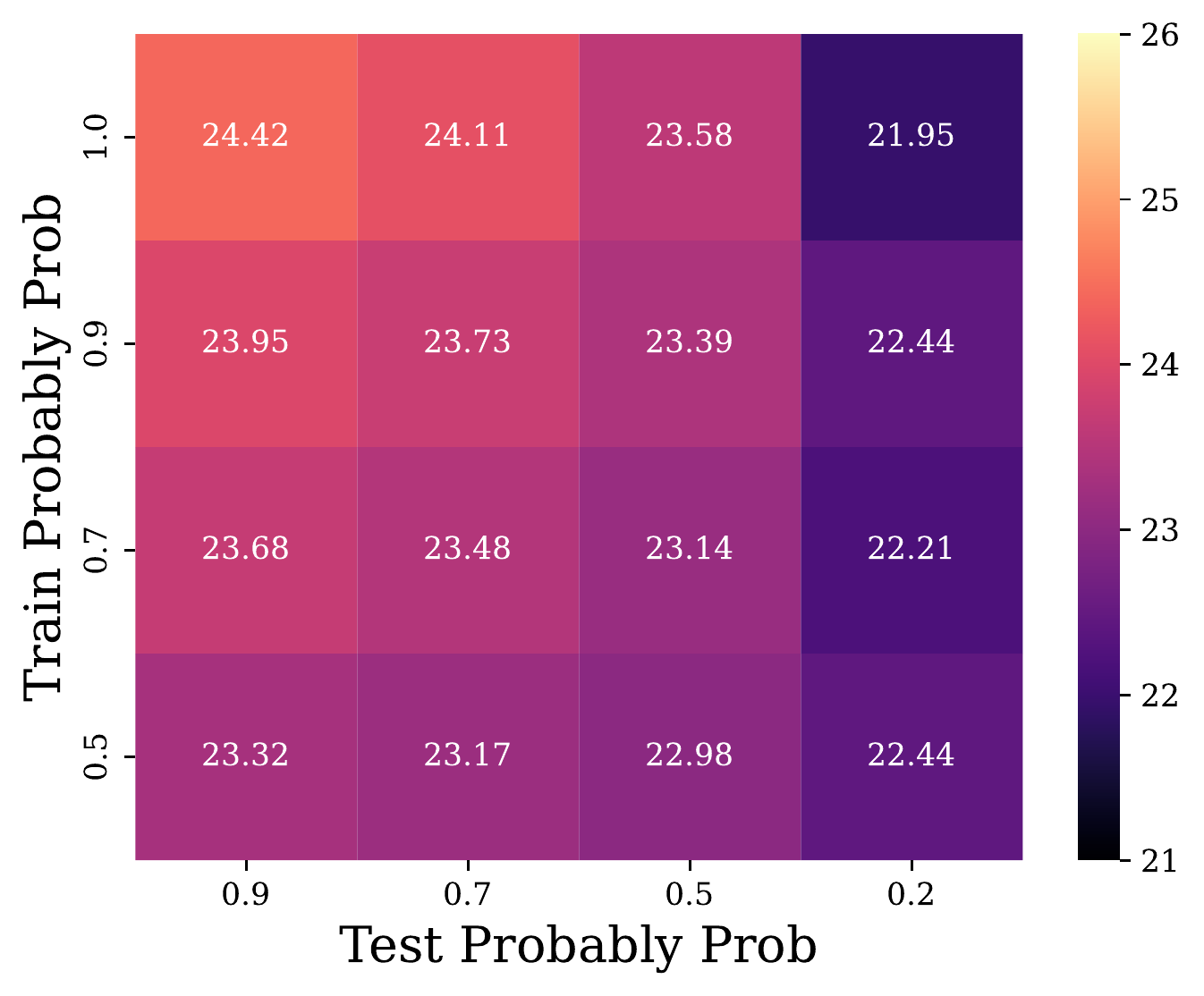}
  \subcaption{Training on instance-level (narrow) uncertain concept annotations.}\label{fig:nosmooth_train_inst_test}
\endminipage\hfill
\minipage{0.32\textwidth}%
  \includegraphics[width=\linewidth]{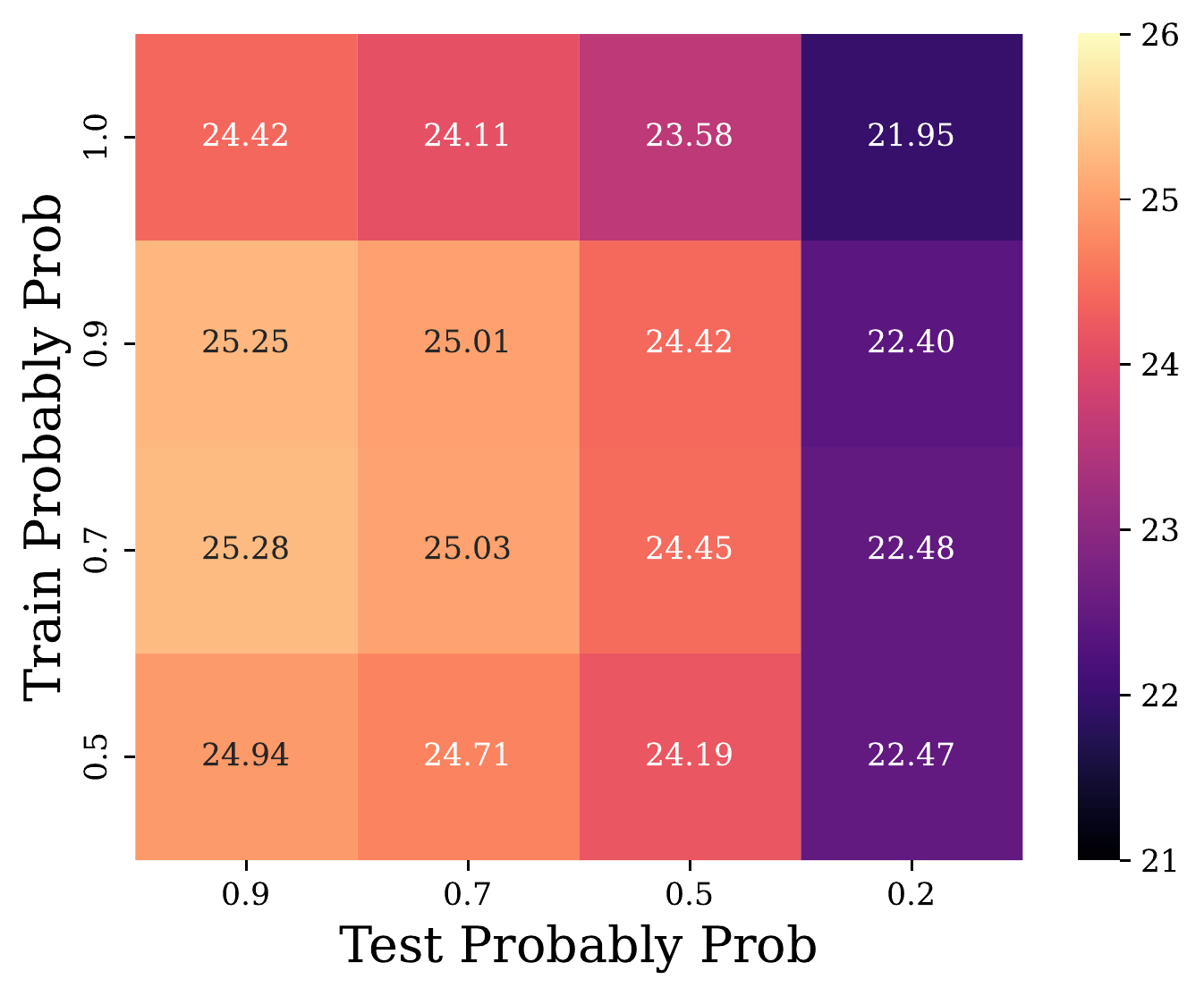}
  \subcaption{Training on population-level (broad) uncertain concept annotations.}\label{fig:agg_train_inst_test}
\endminipage
\caption{Training on uncertain concept labels improves generalization to instance-level (broad) uncertainty at test-time -- the most challenging of the in-the-wild coarse-grained varieties. Heatmap colors depict generalization efficacy operationalized as the AUC between the intervention-accuracy curve. Uncertainty here is expressed by varying the imputed ``Probably'' probability at train and test time; decreasing probability (e.g., left-to-right on the x-axis) corresponds to increasing uncertainty.}
\label{fig:compare_training_labels_cub}
\end{figure*}

Likewise, the question of the form of uncertainty and whether or not to leverage aggregate uncertainty matters at train time. Training on aggregated uncertainty not only performance on similarly population-level uncertainty (see Supplement), but also over softer, potentially noisier individual-level uncertainty -- across gradations of uncertainty (see Figure \ref{fig:compare_training_labels_cub}). Further, whether uncertainty is assumed to be broad or narrow at an individual-level can also impact training label efficacy (see Left and Middle panels of Figure \ref{fig:compare_training_labels_cub}).

\subsubsection{Implications}

While we focus here on \texttt{CUB} -- as the dataset is a highly popular concept benchmark, and therefore necessary to deeply understand -- the elicitation of discrete uncertainty is lightweight and popular in crowdsourcing \citep{uncertainJudgments} (see further investigations with \texttt{CheXpert} in the Supplement); as such, our investigations may be broadly applicable to researchers leveraging elicited discrete uncertainty. The wide impact design choices can have serves as a caution -- if we want safe systems which are robust, we ought to be able to handle the array of intended meaning expressed by humans through discrete uncertainty. Decisions around how to treat discrete uncertainty over concepts persist across train and test time. 

\subsection{Fine-Grained Uncertainty}

Next, we turn to more fine-grained uncertainty. When faced with many options (e.g., multiple different possible colors for the wing, or different gradations of severity in a medical phenotype), a human may have different levels of uncertainty over each option. We now consider this form of categorical uncertainty \textit{explicitly}, rather than inferring from an ambiguous single measure of ``uncertainty.''

However, there is a paucity of datasets available with such richly annotated labelings over concept space. As such, to facilitate this research, \textbf{we build a new platform for uncertainty elicitation over concepts}, which we call \texttt{UElic} and offer a first application of \texttt{UElic} to relabel a subset of \texttt{CUB} with human soft labels over all concepts. \textbf{We release our dataset as \texttt{CUB-S}, replete with nearly 5,000 rich uncertainty-labeled concept groups}. 

In this Section, we begin by introducing our new elicitation interface for rich human uncertainty, and offer several insights into the character of human uncertainty we elicit. We then highlight how concept-based systems crumble under the nuances of the fine-grained uncertainty we elicit. We believe \texttt{CUB-S} can serve as a formative dataset to further study human uncertainty in concept-based models.







\subsubsection{Eliciting Human Uncertainty} 

\begin{figure}[!htb]
    \centering
    \includegraphics[width=0.9\linewidth]{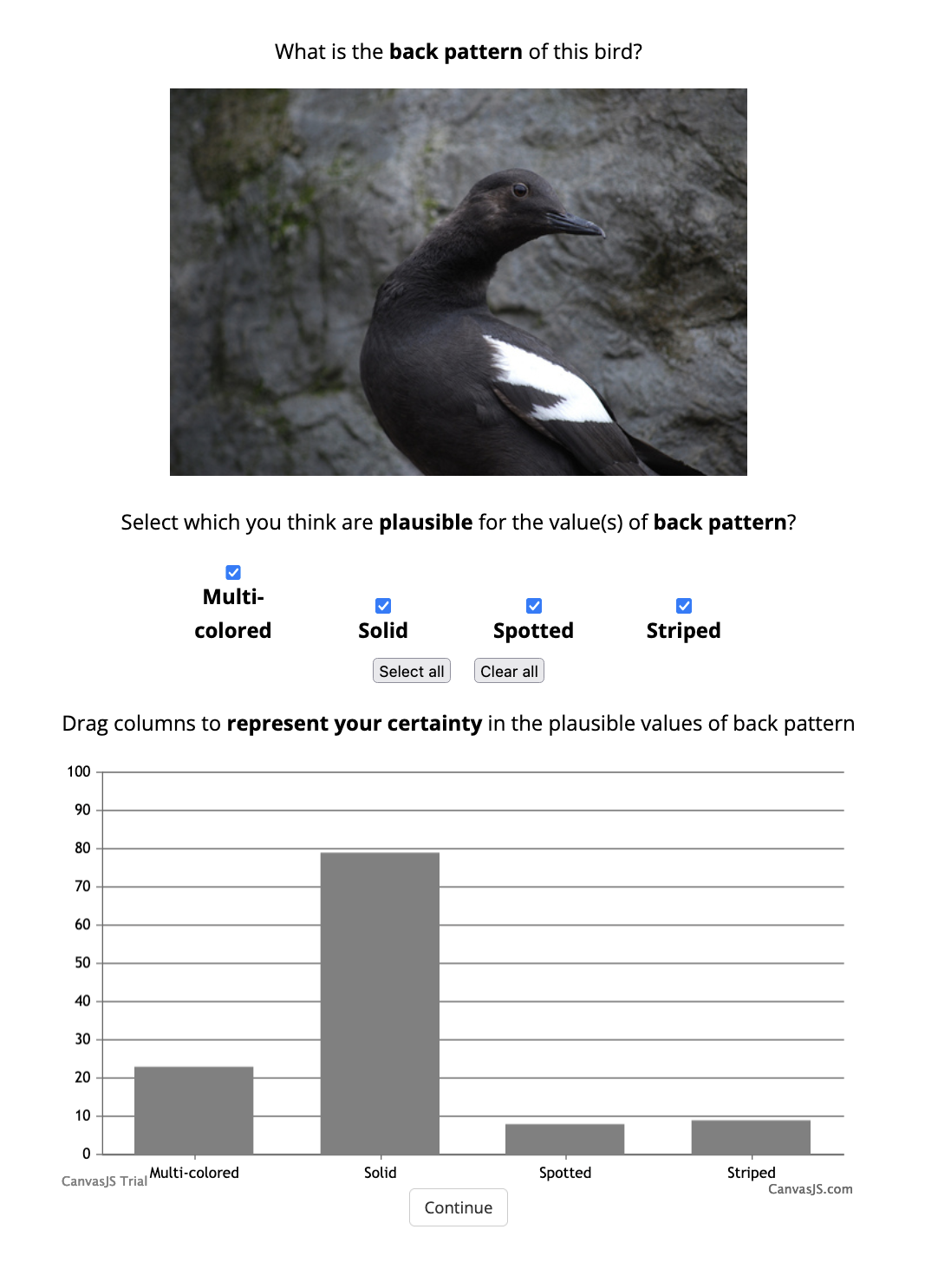}
    \caption{Example screen of \texttt{UElic} for \texttt{CUB}. Participants select the concept attributes they think are plausible, and drag bars to express said uncertainty. Here, the back is not visible; users must be uncertain in their annotation. We empower annotators to \textit{richly} express this belief distribution, in contrast to the original \texttt{CUB} dataset.}
    \label{fig:interface-unc}
\end{figure}

We offer a new platform to streamline the elicitation of human uncertainty. Our interface, \texttt{UElic}, offers a lightweight paradigm for users to express uncertainty. Users are presented with the features of interest (e.g., an image), the concept to be annotated, and all available options. To reduce the cognitive load of expressing uncertainty over \textit{all} options per concept, we request users select only the attributes they think are plausible and express their uncertainty over these attributes by dragging an interactive bar chart to represent their perceived probability, inspired by \citep{goldstein2014lay}. An example interface screen is depicted in Figure \ref{fig:interface-unc}. 

\subsubsection{Collecting CUB-S} We recruit 89 participants from the crowdsourcing platform, Prolific \cite{palan2018prolific}. Participants annotate \textit{all 28 concepts} for two different bird images: totalling \textbf{4984 soft categorical concept group annotations}. Within each soft concept group annotation, participants provide their uncertainty over each of the possible attributes for that concept (e.g., possible wing colors, beak shapes, etc). Stimuli are selected from the \texttt{CUB} test set, and preferentially subsampled to include images which CEMs and CBMs both typically get wrong\footnote{Approximately 50\% of the images shown to participants are those which four different seeds of both CEMs and CBMs got incorrect, rendering them more interesting - and challenging - to study at intervention-time}. Participants are paid at a base rate of \$9/hr, with an optional bonus paid up to \$10/hr to encourage quality predictions; the bonus is applied to all participants.

\subsubsection{Richness in CUB-S} 
\label{sec:cubs_richness}

\begin{figure*}[!htb]
    \centering
    \includegraphics[width=1.0\linewidth]{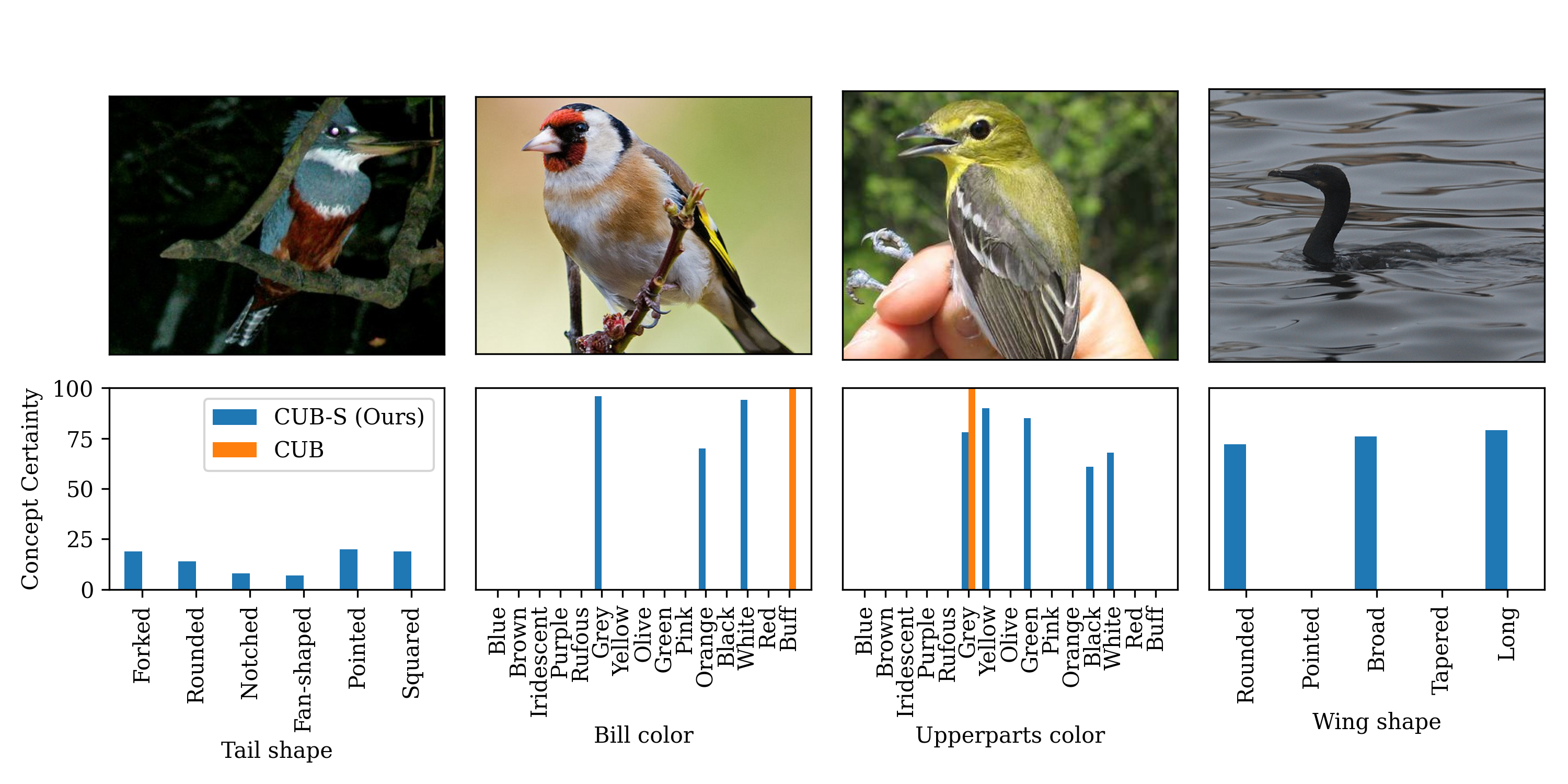}
    \caption{Example soft concept annotations elicited in CUB-S compared to \texttt{CUB} class labels. Far left: well-calibrated annotations for the "tail shape" concept, expressing appropriate uncertainties which sum to 100. Center left: annotations rarely included the obscure "buff" color, even when it was appropriate. Center right: richer annotations for the ``upperparts color'' provide more information than the certain \texttt{CUB} annotations. Far right: uncalibrated uncertainty of the ``wing shape'' concept under occlusion.}
    \label{fig:cubs_concept_examples}
\end{figure*}

Our elicited soft labels demonstrate that humans indeed can starkly depart from a uniform distribution of uncertainty over concepts (see Figure \ref{fig:cubs_concept_examples}). Humans possess rich approximations of uncertainty. Eliciting this uncertainty directly can resolve some of the ambiguities mentioned with discrete uncertainty. 

Further, by tracking which concepts were annotated by particular individuals (information which is not stored in the original \texttt{CUB} annotations), we identify a wide spectrum in the calibration of annotators. This is not entirely unexpected, given different levels of uncertainty calibration in humans broadly \citep{KLAYMAN1999216, keren1987facing}. We use the Expected Calibration Error (ECE) \cite{naeini2015obtaining} as a metric to evaluate the accuracy of annotators when estimating their confidence. Intuitively, the metric is the expected absolute difference between the fraction of correct predictions (accuracy), and the probabilities provided by the annotators (confidence). The ``correct'' concepts for a given bird are determined from the original \texttt{CUB} annotations averaged over all birds of the same species. These ``correct'' concepts are a suitable approximation to ground truth, and are significantly less noisy than the \texttt{CUB-S} annotations; however, we emphasize that they are \textit{not definitive ground truth}. 

\begin{figure}
    \centering
    \includegraphics[width=0.9\linewidth]{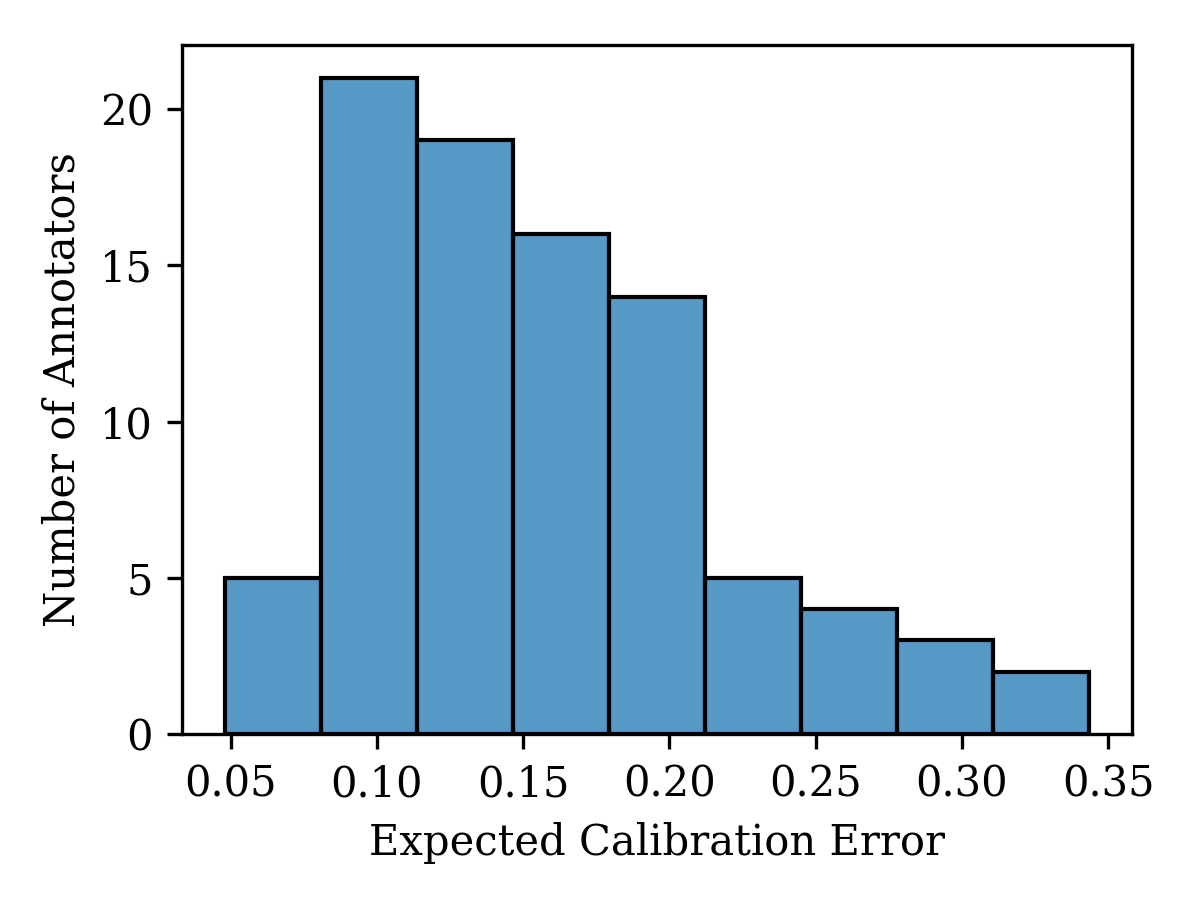}
    \caption{Distribution of Expected Calibration Error for annotators in \texttt{CUB-S}. The positive skew shows most annotators are well-calibrated, with a few who are very poorly calibrated.}
    \label{fig:cubs_miscalib}
\end{figure}

Figure \ref{fig:cubs_miscalib} shows that the majority of annotators are reasonably calibrated, although this value is positively skewed by the large number of (correct) zero probabilities provided for rare concepts (such as the color ``purple''). There are some annotators who are poorly calibrated and it mitigating this issue remains an open question. Some calibration ``error'' is a result of the additional richness in the CUB-S annotations not present in CUB. However, there are also genuine annotation errors which we observe when manually checking the annotations. Illustrative examples comparing soft CUB-S annotations to hard \texttt{CUB} annotations are shown in Figure \ref{fig:cubs_concept_examples}. Humans who intervene at test time will also suffer from calibration errors, challenging the common assumption that human experts are perfect ``oracles''.

\begin{figure}
    \centering
    \includegraphics[width=0.9\linewidth]{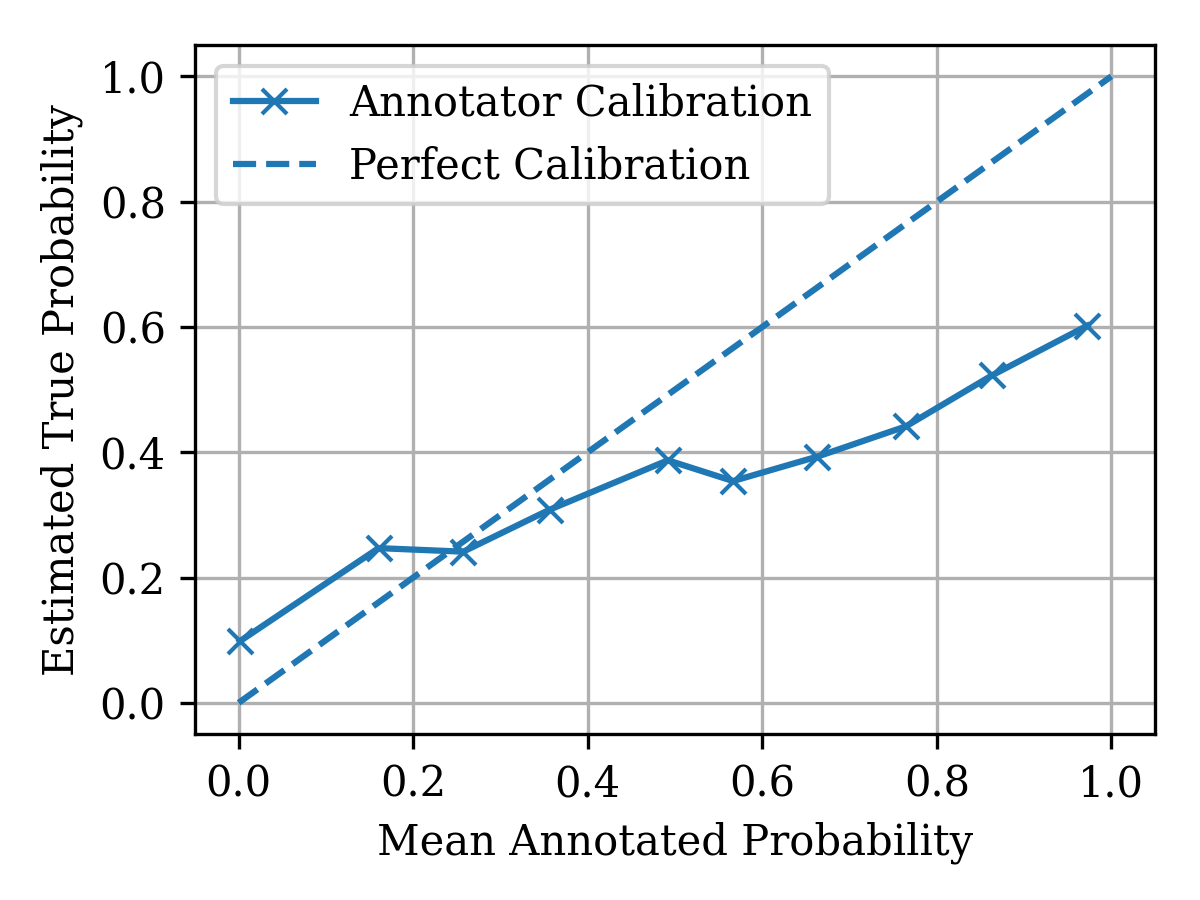}
    \caption{Calibration curve for CUB-S annotators, showing consistent underestimates of small probabilities and overestimates of large probabilities.}
    \label{fig:cubs_calibration_curve}
\end{figure}

On average, we observe that annotators consistently underestimate small probabilities but overestimate large probabilities (Figure \ref{fig:cubs_calibration_curve}). When several concepts are possible, it is likely that annotators attempt to reduce their cognitive load by only selecting a few to have a nonzero probability. Conversely, when a concept is highly probable, annotators may incorrectly round an annotation to 100 (i.e. absolute certainty). Figure \ref{fig:cubs_annotation_hist} shows that 0 and 100 are the most popular uncertain annotation values, due to the presence of these two effects. We emphasize that some errors are predictable, and thus have the potential to be corrected when training an uncertainty-aware model.

It is unclear whether the poor calibration is a result of our interface, or simply an unavoidable issue when eliciting uncertainties in a crowdsourcing setting; humans have limited cognitive resources at any given time -- they may not be willing to endorse several related concepts (e.g., orange and red), while providing detailed uncertainty over each. However, the fact that we \textit{do} encounter such challenges is an important consideration in the deployment of systems in which \textit{receive} such uncertainty estimates. It is essential that \textbf{systems be robust to these nuances and peculiarities in elicited human uncertainty, or else they may fail at deployment time}. 


\subsubsection{Intervening at Test-Time with \texttt{CUB-S}} 

We next apply the same computational investigations as in our prior experiments to \texttt{CUB-S}; the same Experimental Set-Up is applied, now, only varying the labels used at test time. We use models trained on population-level broad uncertainty derived from coarse-grained \texttt{CUB} as in the prior section. 

\begin{figure}[!htb]
    \centering
    \includegraphics[width=0.9\linewidth]{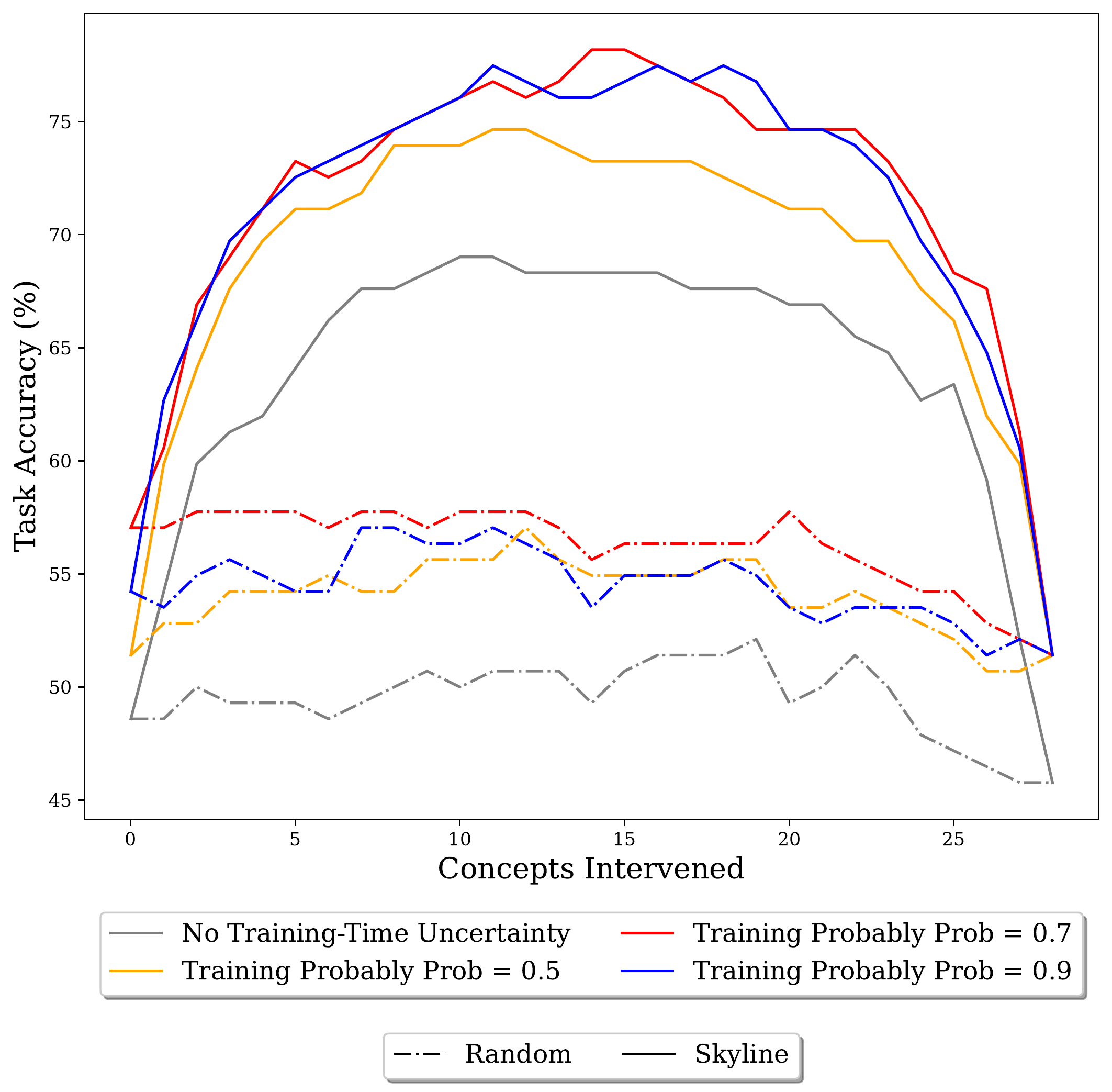}
    \caption{CEMs struggle to handle real, human uncertainty. While Skyline is still able to pick up \textit{some} signal to leverage in the data, not all incorporated concept annotations help model performance: some may hurt. Using models trained on human uncertainty information may mitigate some of the drop.}
    \label{fig:cubs_policies}
\end{figure}

We find in Figure \ref{fig:cubs_policies} that the richness of \texttt{CUB-S} poses a substantial challenge for concept-based models. While we find that using models trained on the coarse-grained uncertainty in \texttt{CUB} can mitigate some of the failures under test-time uncertainty, they are not a perfect salve.

\begin{figure}[!htb]
    \centering
    \includegraphics[width=0.9\linewidth]{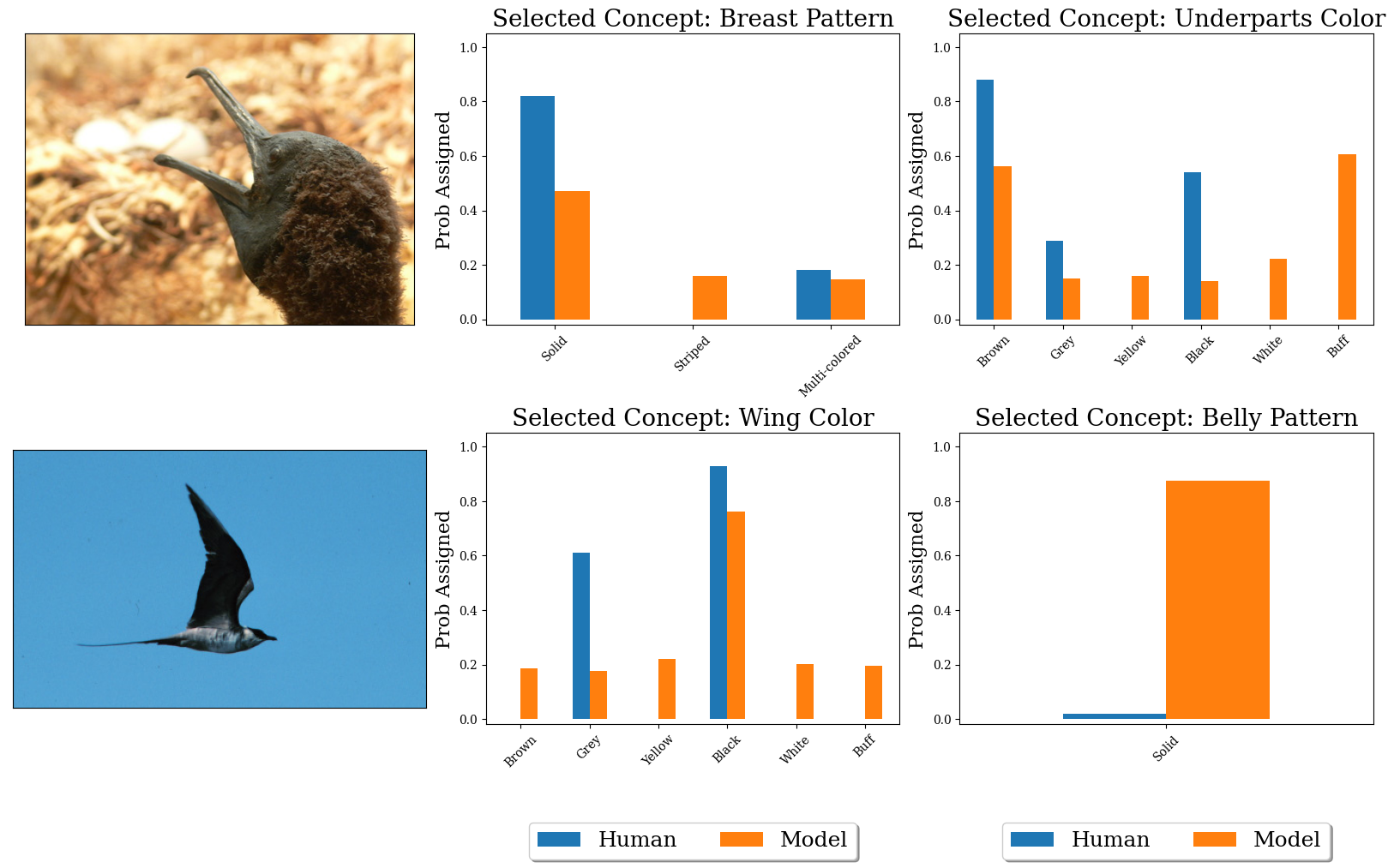}
    \caption{Model versus human distributions over concepts at the time of selection by Skyline. The first column of distributions are selections which boosted the model's classification (from incorrect to correct); humans' uncertainty was helpful to intervene with. The second column of distributions depicts the human uncertainty at intervention time which \textit{hurt} model performance (the classification went from correct to incorrect). Model trained on uncertain concepts (``Probably'' probability = 0.7).}
    \label{fig:model_v_human_uncs}
\end{figure}

The development of better mitigation strategies to handle the nuances of in-the-wild categorical uncertainty over concepts is exciting ground for future work. We observe that some concepts are preferable to elicit interventions over; sometimes human uncertainty is helpful, other times it \textit{harms} model performance (see Figure \ref{fig:model_v_human_uncs}). Further, these differences persist across methods of training the models (i.e., the level of uncertainty in the training data, see Supplement), underscoring the need for adaptive, query procedures personalized to individual- and model uncertainty. We argue multi-disciplinary methodological advances to handle in-the-wild, rich human uncertainty over concept annotations are essential. 




\subsubsection{Implications}

Humans interpret and reason in the world with richly structured uncertainty. Our CUB-S elicitation demonstrates this richness. However, we find that concept-based systems struggle to handle this level of richness. Given humans \textit{are capable and do} express fine-grained uncertainty, it is sensible that our systems ought to be equipped to handle the nuances of in-the-wild uncertainty.

\section{Open Challenges}

We emphasize the importance of considering human uncertainty in concept-based models, and the need for richer datasets of human uncertainty to study these challenges. \texttt{CUB-S} is a promising initial playground to study the nuances faced with real human uncertainty\footnote{All code and data will be hosted at our \href{https://github.com/collinskatie/uncertainty-concepts}{repository}.}. Our work raises several open challenges.


\subsection{Complementarity of Human and Machine Uncertainty}

Considering human uncertainty in interventions opens up exciting opportunities in the study of human-machine complementarity \cite{branson20questions, wilder2020learning, bondi2022role, steyvers2022bayesian}. 
When we break the assumption that humans are confident oracles, it becomes especially important to consider whether cases which are hard for the model to annotate are also those that a human struggles with; in that case, selecting such a concept is not ideal. Learning models and intervention policies which can complement humans' strengths and weaknesses, accounting for their expertise and confidence, are promising grounds for further study with \texttt{CUB-S} and beyond. We see that varied models may prefer different forms of uncertainty (see Supplement); further, even though we see that Random interventions fail disastrously, there is a signal that Skyline picks up on the uncertain concepts -- how can we predict where and when to ask people for their uncertainty? And when we do receive their uncertainty, it is not immediately apparent whether we \textit{should} take the human intervention as ``truth.'' As we demonstrate, real humans are \textit{not} consistent oracles -- and in some settings (e.g., occlusion), \textit{no} human may be an oracle, even if they are an expert. Models which can learn whether or not to trust human interventions, e.g. \citep{dvijotham2022enhancing, conceptSidecar}, are promising grounds for future study.


\subsection{Treating Human (Mis)Calibration}

A core factor in whether or not to trust a \textit{human's} intervention, and determinant of which concepts to query, may depend on the expected calibration of the user. We observe wide variation in individuals' level of calibration in their uncertainty expression, a finding that resonates with a wealth of cognitive science literature \citep{lichtenstein1977calibration, KLAYMAN1999216, keren1991calibration, keren1987facing, kahneman1996reality, sharot2011optimism, uncertainJudgments}. However, we emphasize that calibration need not be a turn-off from collecting uncertainty in the first place; not only are some humans highly calibrated -- but forcing someone to express certainty when they are not (and when it is not possible to be certain; e.g., occlusion), we argue may be worse. Future work for post-hoc calibration in a \textit{few-shot} manner, i.e., from limited individual-level user data, provided in an online fashion, is further promising ground for new methodological advances. Additionally, we encourage further experimentation with \texttt{UElic} to encourage better calibration from humans -- perhaps through the use of a carefully designed teaching curriculum \citep{keren1991calibration, keren1990cognitive}. We see calibration as an exciting nexus for a multi-disciplinary study spanning ML, cognitive science, UX design, and psychology.


\subsection{Scaling Uncertainty Elicitation}

Further, we recognize that the annotation of large-scale datasets with human uncertainty may be practically challenging. It is costly to elicit human uncertainty: annotators take substantially more time \citep{selfCiteSoftLabel}. There is a need for more scalable elicitation techniques, and better simulators of human uncertainty to permit the study of softness at train time. We observe substantial differences in model performance depending on the form of uncertainty used; more data is needed to further characterize these differences and determine when one form of uncertainty is better to elicit than another, such that when we deploy systems in the world -- they can handle a variety of forms of uncertainty expression.

\section{Conclusion}

We highlight the importance of considering human uncertainty in concept-based models in order to improve reliable performance for safe applications in deployment across society. Humans in the real world are not certain oracles. We make mistakes and may be unsure. Even though humans may be miscalibrated in their uncertainty, we believe the study of tools to elicit and work with human uncertainty has great potential to improve human-in-the-loop systems. Through a mixture of simulated and in-the-wild experiments with uncertainty, we demonstrate failure modes of popular concept-based systems to handle both coarse- and fine-grained uncertain feedback. We offer a new interface, \texttt{UElic}, and a new challenge dataset, \texttt{CUB-S}, to support further study into human uncertainty in interventions. Modeling human uncertainty at train- and test-time has the potential to greatly improve the reliability and trustworthiness of concept-based models when deployed safely in the wild.

\section*{Acknowledgments}

We thank Yanzhi Chen, Isaac Reid, Kris Jensen, Richard Turner, Carl Henrik Ek, and Josh Tenenbaum for helpful conversations. We also thank Steve Branson, Catherine Wah, Kushal Chauhan, and Rishabh Tiwari for helpful clarifications on their work. Thank you to the participants from Prolific who took part in our annotation.

KMC gratefully acknowledges support from the Marshall Commission and the Cambridge Trust. MEZ acknowledges support from the Gates Cambridge Trust via the Gates Cambridge Scholarship. NR is supported by a Churchill Scholarship. UB  acknowledges  support  from  DeepMind  and  the  Leverhulme Trust  via  the  Leverhulme  Centre  for  the  Future  of  Intelligence  (CFI),  and  from  the  Mozilla  Foundation.  MJ is supported by the EPSRC grant EP/T019603/1. AW  acknowledges  support  from  a  Turing  AI  Fellowship  under grant  EP/V025279/1,  The  Alan  Turing  Institute,  and  the Leverhulme Trust via CFI. IS is supported by an NSERC fellowship (567554-2022).

\bibliography{main}

\newpage

\section*{Supplement}

\subsection*{Constructing \texttt{UMNIST}}

\begin{figure}[!htb]
    \centering
    \includegraphics[width=1.0\linewidth]{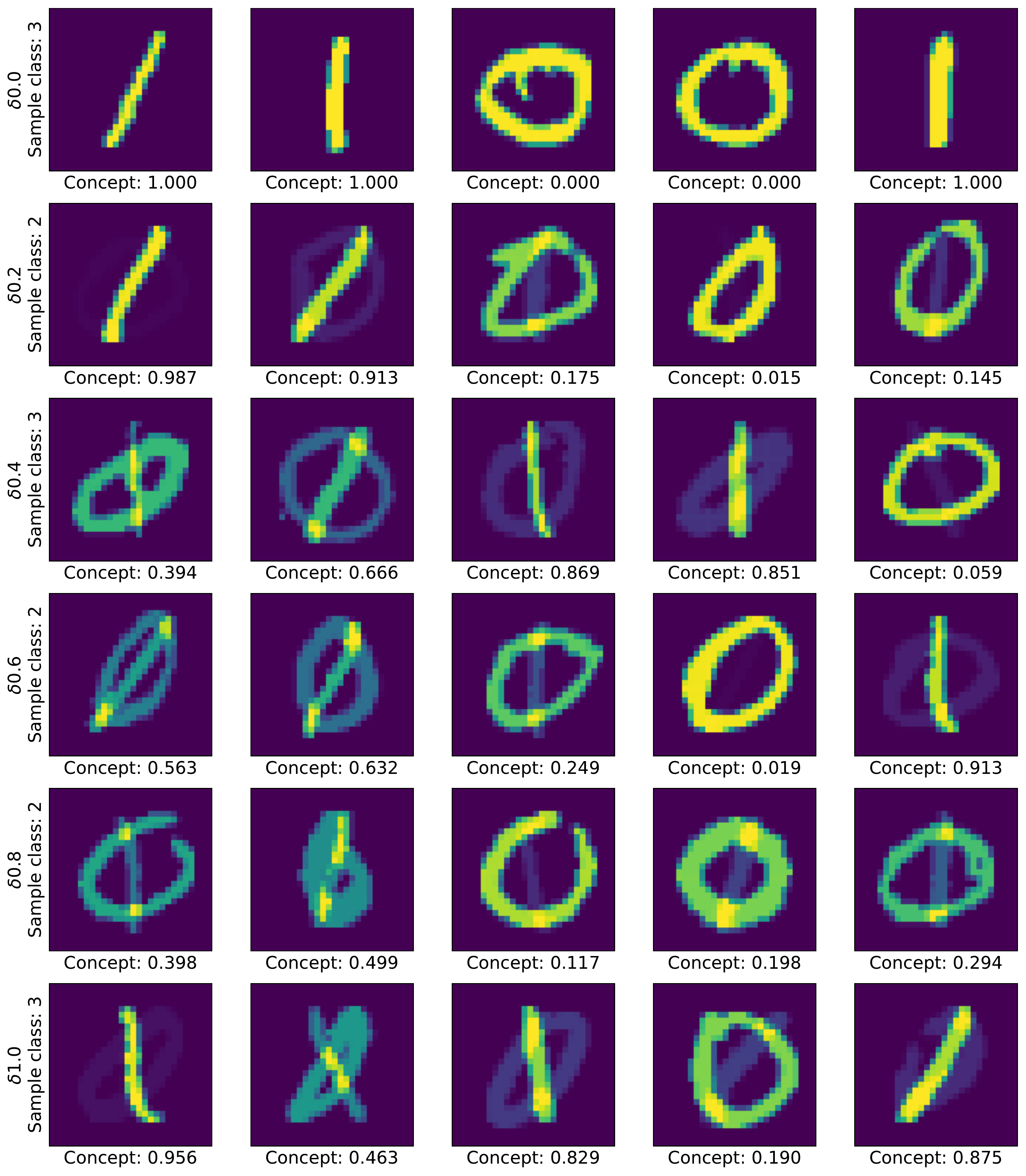}
    \caption{Example datapoints in \texttt{UMNIST} as we vary the value of $\delta$ (rows). Each row represents a single sample, with each column representing one of the $p=5$ digits forming that sample. We include each concept's annotation, as well as the datapoint's label, underneath each digit and to the left of each datapoint, respectively.}
    \label{fig:umnist_example_data}
\end{figure}

We provide further clarity on how we constructed \texttt{UMNIST}. Each sample of the \texttt{UMNIST} dataset is formed by $p$ $28\times28$ grey-scale images of handwritten zeros or ones, given as a normalized sample with shape $\mathbf{x} \in [0, 1]^{28 \times 28 \times p}$. We annotate each sample with $p$ binary concept annotations $\mathbf{c} \in \{0, 1\}^p$, where $c_i$ indicates whether the $i$-th image is a one or a zero, and a task label $y \in \{0, \cdots, p\}$ corresponding to the number of ones in its digits, i.e., $y = \sum_i c_i$. To introduce uncertainty in this dataset's samples and concepts, we update concept $c_i$ corresponding to the $i$-th image as follows:
\[
    c_i := \begin{cases}
        \text{Randomly sample from Unif}(0, \delta) & \text{if i-th digit is 0}\\
        \text{Randomly sample from Unif}(1 - \delta, 1) & \text{if i-th digit is 1}\\
    \end{cases}
\]
where $\delta \in [0, 1]$ is a user-provided hyperparameter controlling the amount of dataset uncertainty. Furthermore, in order for this concept annotation uncertainty to be reflected as part of the input digits $\mathbf{x}$, we mix a concept's corresponding digit, akin to~\citet{mixup}, with a randomly selected MNIST training example of the opposite digit using $c_i$ as the mixing ratio. In other words, after generating a sample's uncertain concept annotations $\mathbf{c}$ we update its $i$-th input digit $\mathbf{x}_{(:, :, i)}$ as follows:
\[
    \mathbf{x}_{(:, :, i)} := \begin{cases}
        (1 - c_i) \mathbf{x}_{(:, :, i)} + c_i \mathbf{z} \text{ with } \mathbf{z} \sim p_\text{M}(\mathbf{x} | y = 1) & \text{if } \mathbf{x}_{(:, :, i)} \text{ is 0}\\
        c_i \mathbf{x}_{(:, :, i)} + (1 - c_i) \mathbf{z} \text{ with } \mathbf{z} \sim p_\text{M}(\mathbf{x} | y = 0) & \text{if } \mathbf{x}_{(:, :, i)} \text{ is 1}\\
    \end{cases}
\]
where $p_\text{M}(\mathbf{x} | y)$ is the empirical training distribution of MNIST samples whose label is $y$. For this paper, we focus on using only $p = 10$ digits per sample. See Figure \ref{fig:umnist_example_data} for some examples of this dataset as we vary $\delta$. 

\subsection*{Computational experiment details} 

We next include additional details on how models were trained and run on the various probe datasets, as well as the intervention methods considered.

\subsubsection*{Training Details for \texttt{UMNIST} Experiments}
For all \texttt{UMNIST} experiments, we train both CBMs and CEMs using a concept extractor whose architecture consisted of four 3-by-3 convolutional layers with filters $\{5, 10, 20, 40\}$ followed by a linear layer with $20$ activations and an output layer with $pm$ output activations, where $m$ is the embedding size used for CEM (one can think of CBM as having $m=1$). In practice, we set $m$ to $8$ following the recommendations from~\citet{cem22}. Between all non-output layers, we include leaky-ReLU nonlinear activations and we apply batch normalization after each nonlinearity that follows a convolutional layer. Similarly, for both CEMs and CBMs, we use a simple ReLU two-layer MLP as its concept-to-label map with layers sizes $\{20, p\}$ and train each end-to-end CBM/CEM by weighting the concept loss as much as the task loss (i.e., the joint training hyperparameter $\alpha$ was set to $\alpha = 1$ for both methods). Finally, to avoid each model learning to simply predict the most common class to minimize its error, we weight each sample's task loss according to the empirical label distribution of its corresponding label to encourage our models.

All models are trained by sampling a total of $20,000$ training \texttt{UMNIST} samples, of which 20\% were used as a validation set, and tested by sampling $5,000$ \texttt{UMINST} testing samples from MNIST's testing set (so no digit in the testing set is ever used to construct \texttt{UMNIST}'s training set). We train all models using a standard Adam \cite{kingma2014adam} optimizer with a learning rate $10^{-3}$ and a batch size of $256$ for a maximum of $50$ epochs, stopping earlier if the validation loss has not improved for $15$ epochs. For each method in \texttt{UMNIST}, we run 5 models from different seeds.

\subsubsection*{Training Details for \texttt{CheXpert} Experiments} 
For the \texttt{CheXpert} dataset \citep{irvin2019chexpert}, we train all models for 25 epochs, subsampling the dataset to use only 25\% of the training dataset when training due to the large size of the dataset. 
Because the test split for \texttt{CheXpert} does not have the ``uncertain'' concept label, we perform an 80-10-10 split of the train split into the train, validation, and test folds. Results for \texttt{CheXpert} are averaged over 5 trials, and we use a learning rate of 0.001 across all trials. 

\subsubsection*{Training Details for CUB-Based Experiments} Models trained on \texttt{CUB} followed the same training settings as~\citet{cem22}; we employ a single model run for each seed due to computational complexity.

\subsubsection*{Details on Intervention Policies} The interaction policies we consider in this work (Random and Skyline) consider the setting where a user can be queried to intervene, or edit, a single concept (e.g., wing color) at a time. \textit{Skyline} assumes access to the true label $y$ and how the human would intervene (e.g., assumes access to the \texttt{CUB-S} elicited soft concept annotations), and ``tests'' intervening with each of the remaining concepts to see which yields the highest predicted probability of the model on the true label. In that way, this mimics an ``Oracle'' policy, which can greedily select the best of the available next concept interventions, following \citeauthor{chauhan2022interactive}. However, the assumption of knowing the humans' interventions in advance, and the true label, are not realistic (and defeat the purpose of an intervention policy) in practice; hence, this method is meant to capture the ``best possible'' amount of information that can be gleaned by a single-step direct intervention policy alone. \textit{Random} simply selects the next concept to query by randomly choosing amongst the available concepts which have not yet been queried.

\subsection*{Additional Results on Concept-Incomplete Variant of \texttt{UMNIST}}
As discussed by~\citet{cem22}, CBMs have a significant failure mode when the set of concept annotations available at training time is not fully predictive, or complete, with respect to the task of interest. Similarly to our \texttt{UMNIST} experiments summarized in Figure~\ref{fig:umnist_res}, this section explores how test-time uncertainty affects CBMs and CEMs when the dataset we are working with does not have a complete set of concept annotations. For this, we use our defined \texttt{UMNIST} dataset but only provide $50\%$ of its concept annotations at training time. We train a CBM and CEM using the same configuration and architecture as that described for our \texttt{UMNIST} experiments, with the exception that the concept weight loss $\alpha$ was changed to $0.1$. We apply such a change to improve CBM's performance, as otherwise, it was unable to achieve a moderately high task accuracy.

\begin{figure}[!htb]
    \centering
    \includegraphics[width=\linewidth]{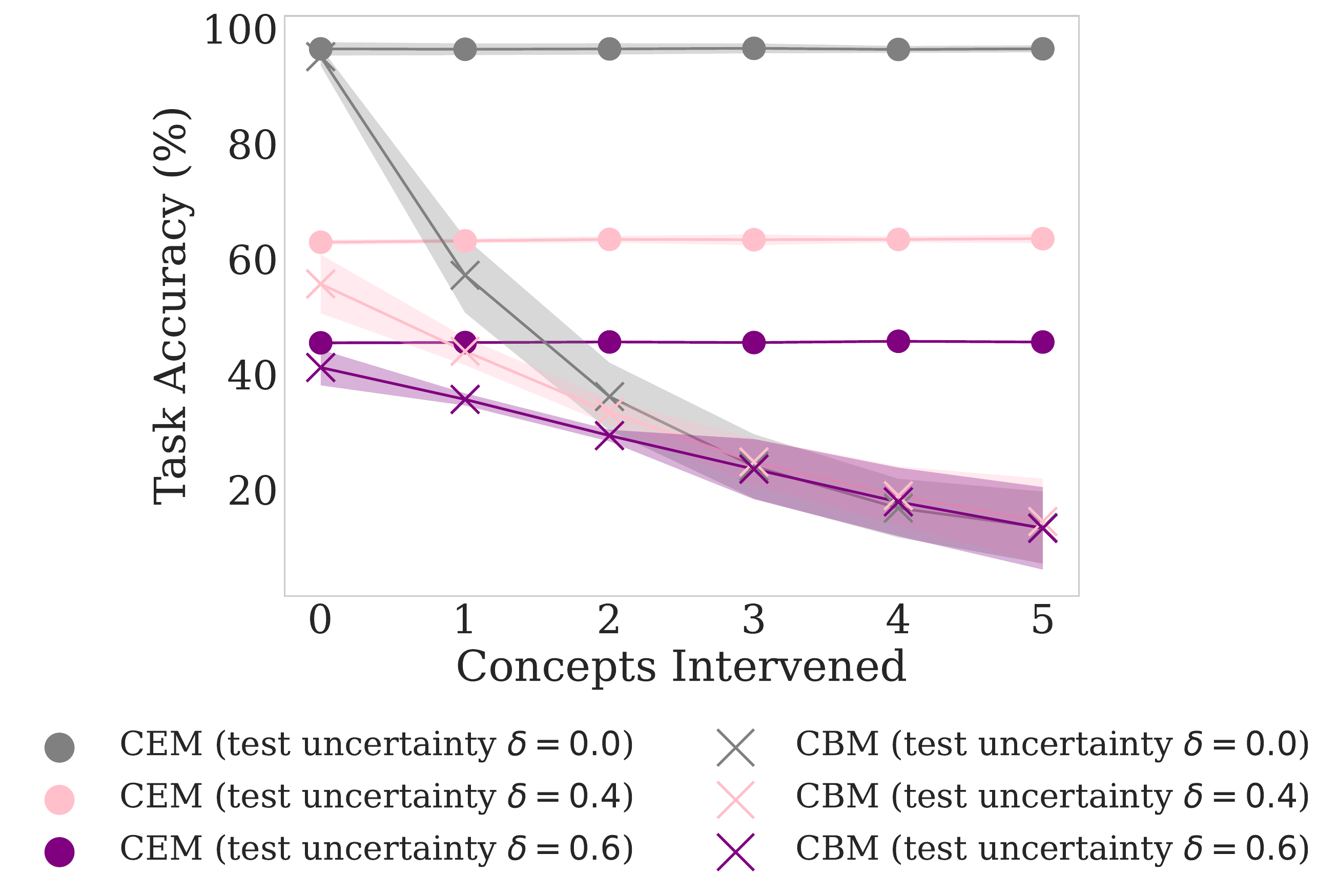}
    
    \caption{Mean test accuracies of random interventions on CBMs and CEMs, and standard errors across 5 different random initializations, as we increase the number of concepts we intervene on. These models are trained on a variant of \texttt{UMNIST} where we only provide $50\%$ of its concepts at training time. 
    }
    \label{fig:umnist_res_incomplete}
\end{figure}

Our results in Figure~\ref{fig:umnist_res_incomplete} demonstrate that both CBMs and CEMs significantly drop their performance when test-time uncertainty increases (as we saw in Figure~\ref{fig:umnist_res} before). Nevertheless, in contrast with Figure~\ref{fig:umnist_res}, we see that interventions in CBMs actually decrease their test accuracy, with uncertainty exacerbating this effect even further. Therefore, these experiments suggest that in concept-incomplete setups, which tend to be what we would expect in real-world datasets given the cost of acquiring all possible concept annotations, CEMs are relatively safer to use regardless of the user's uncertainty at intervention time.

\subsection*{Additional \texttt{CheXpert} Investigations}

\begin{figure}[!htb]
    \centering
    \includegraphics[width=1.0\linewidth]{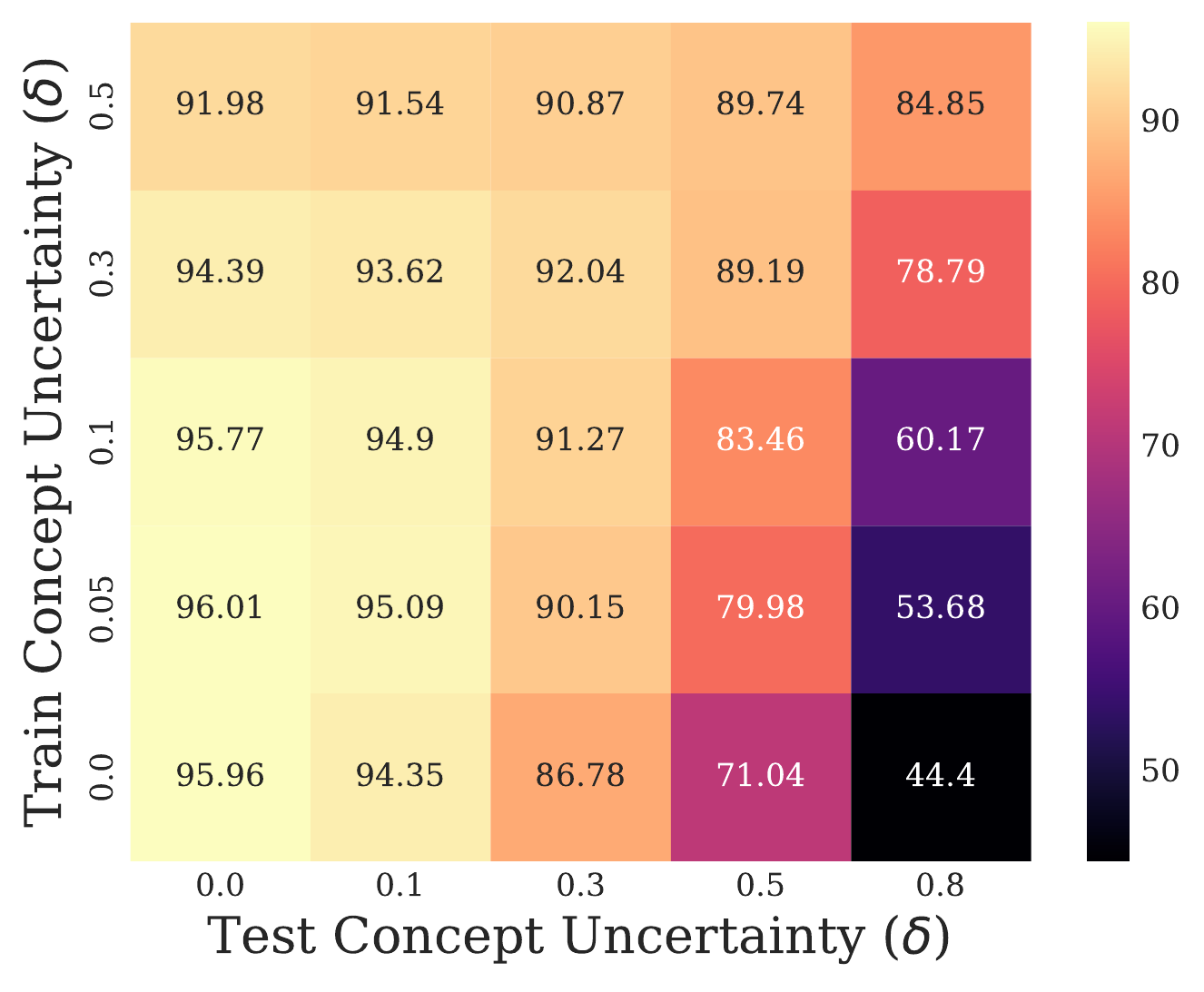}
    \caption{Comparing CEMs trained and tested on differing levels of uncertainty in \texttt{CheXpert}. Heatmap colors depict AUC of the different variants.}
    \label{fig:chexpert_sim_unc}
\end{figure}

We include training with simulated uncertainty in Figure \ref{fig:chexpert_sim_unc}.

\begin{figure}[!htb]
    \centering 
    \includegraphics[width=0.9\linewidth]{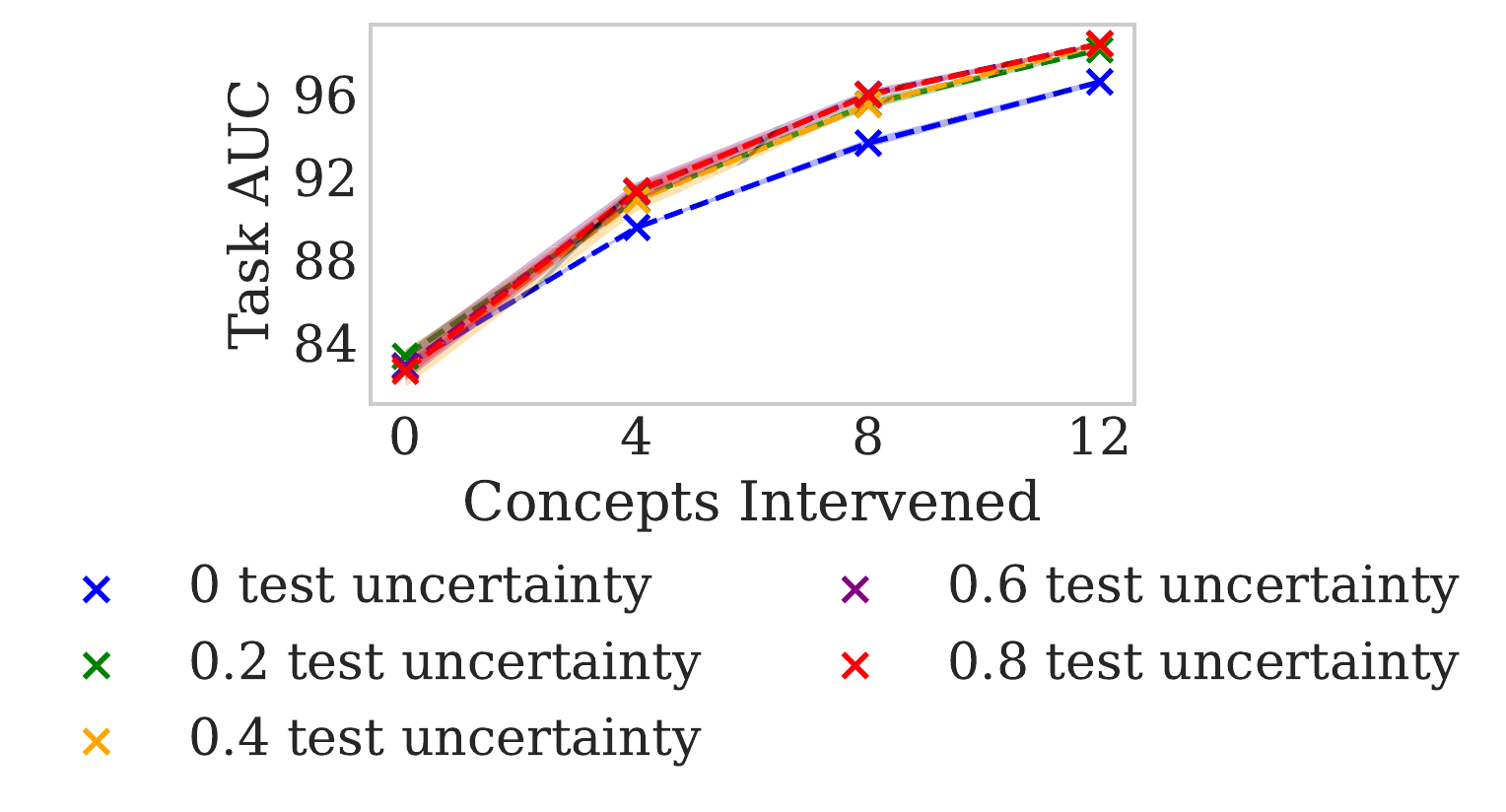}
    \caption{Test-time uncertainty values have a large impact on intervention performance, when using random concept interventions. Setting it to $0$ prevents models from differentiating between negative concepts and uncertain concepts, leading to a decrease in performance. However, setting it to non-zero values allows models to pick up on this difference and improve intervention performance. } 
    \label{fig:chexpert_test} 
\end{figure}

\begin{figure}[!htb]
    \centering
    \includegraphics[width=1.0\linewidth]{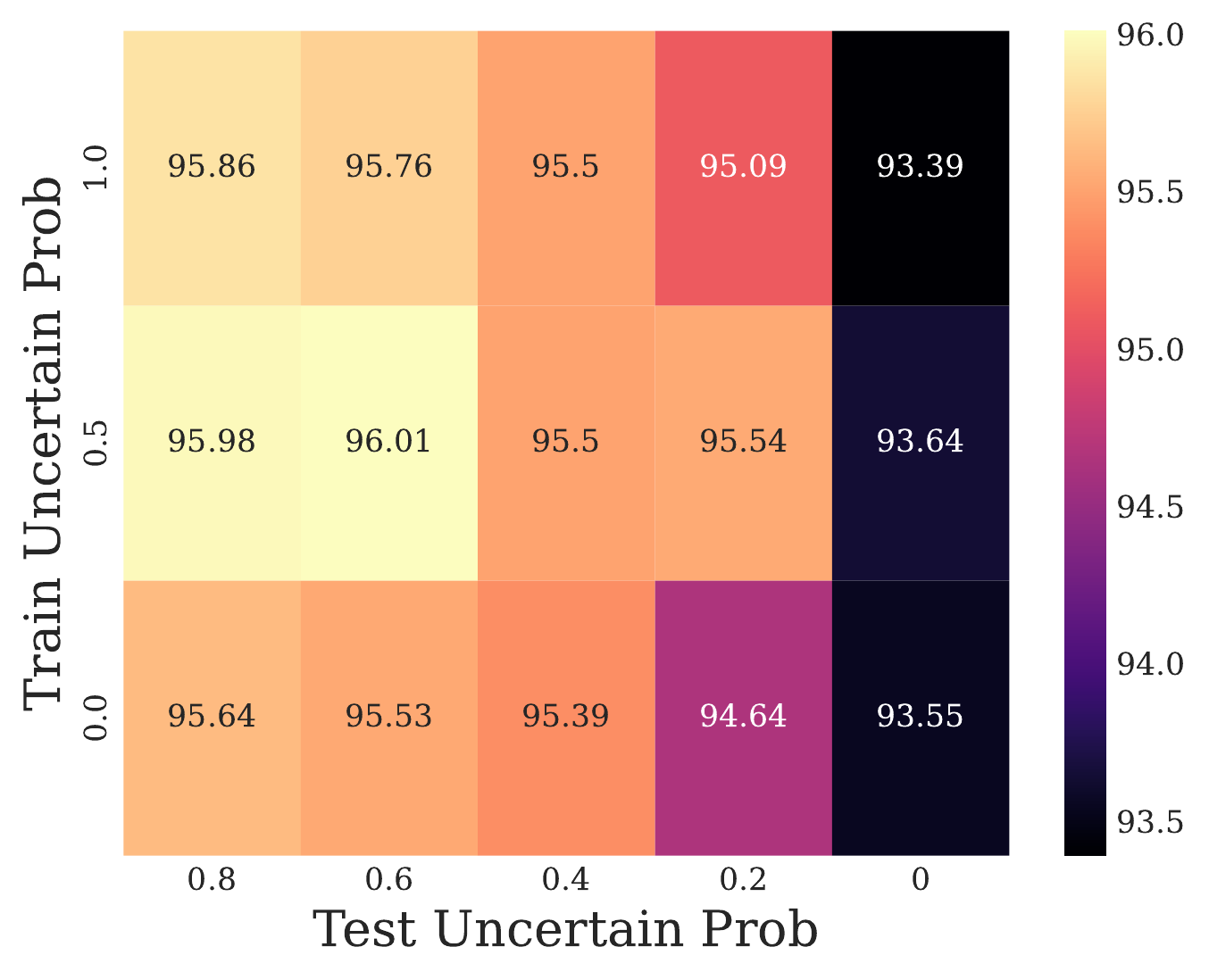}
    \caption{Intervention performance (i.e., random interventions) when using 8 out of 13 concepts to intervene across training and testing uncertainty values. Test uncertainty values have a much larger impact than train uncertainty values, and in general, training with uncertainty seems to have little impact on test-time uncertainty performance.}
    \label{fig:heatmap_chexpert}
\end{figure}

We also explore the original uncertain annotations in \texttt{CheXpert} \citep{irvin2019chexpert}, which contains concept annotations from chest x-rays. The dataset is marked with four labels: positive, negative, unknown, and uncertain. For our experiments, we vary the value taken by uncertain labels, both at train and test time, and investigate its impact on intervention performance. In Figures and \ref{fig:chexpert_test} and \ref{fig:heatmap_chexpert}, we find that test-time uncertainty improves intervention performance, while train-time uncertainty has minimal impact. 
This is partially because of the sparsity of uncertain labels in the dataset; only 5\% of annotations are marked as uncertain, capping the total effect of train-time uncertainty. 
For test-time uncertainty, wen find that non-zero values improve intervention accuracy, because models are able to distinguish between ``uncertain'' labels and ``negative'' labels. 

\subsection*{Additional Details on CUB-S}

\begin{figure}[!htb]
    \centering
    \includegraphics[width=0.9\linewidth]{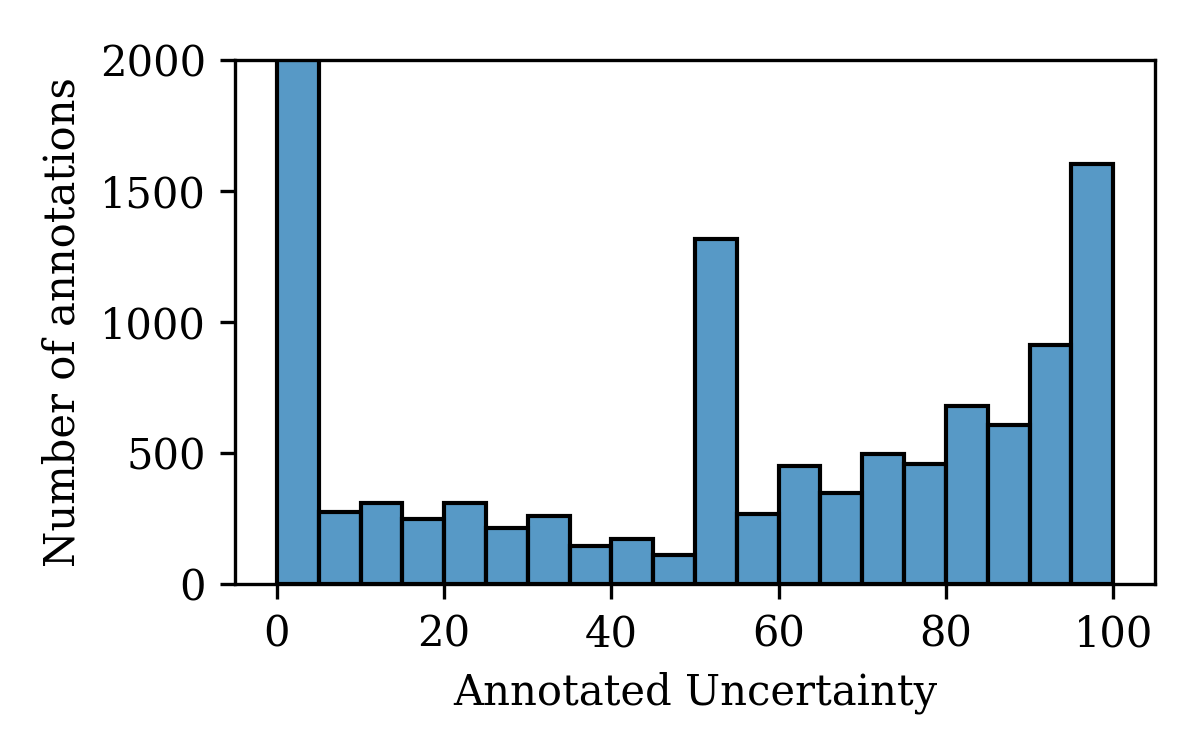}
    \caption{Distribution of uncertainty values for all annotations in \texttt{CUB-S}. Annotators favor certain annotations (0 or 100) and the default value of 50 provided by the interface.}
    \label{fig:cubs_annotation_hist}
\end{figure}

We next include further qualitative observations into the CUB-S soft concept labels collected. We observe in Figure \ref{fig:cubs_annotation_hist} that distribution of provided uncertain annotations is highly irregular, with heavy tails at 0 and 100, and a peak at 50. We hypothesize that heavy tails may be explained by humans rounding values to reduce their cognitive load; \citeauthor{selfCiteSoftLabel} found similar rounding effects in free-form uncertainty expression. 50 is the default value provided by the interface, likely underlying the large number of annotations at 50. This suggests there is scope for improving the interface to extract a more accurate distribution of uncertainties, potentially striking a more naunced balance in granularity of information elicited, e.g., \citep{efficientElic}.

As observed in Section \ref{sec:cubs_richness}, the calibration of individual annotators varies significantly. Figure \ref{fig:cubs_annotation_hist} shows that most annotators consistently assign approximately 100 probability mass for each concept, as one would expect. However, the distribution is positively skewed, with a significant number of annotators consistently over-assigning probability mass acoss the concept groups (for any individual concept, the annotator can endorse at most 100 ``probability units''). This is partly explained by concept groups where more than one concept is relevant (such as color), although it is also likely that annotators are overestimating their confidence. 

\begin{figure}[!htb]
    \centering
    \includegraphics[width=0.9\linewidth]{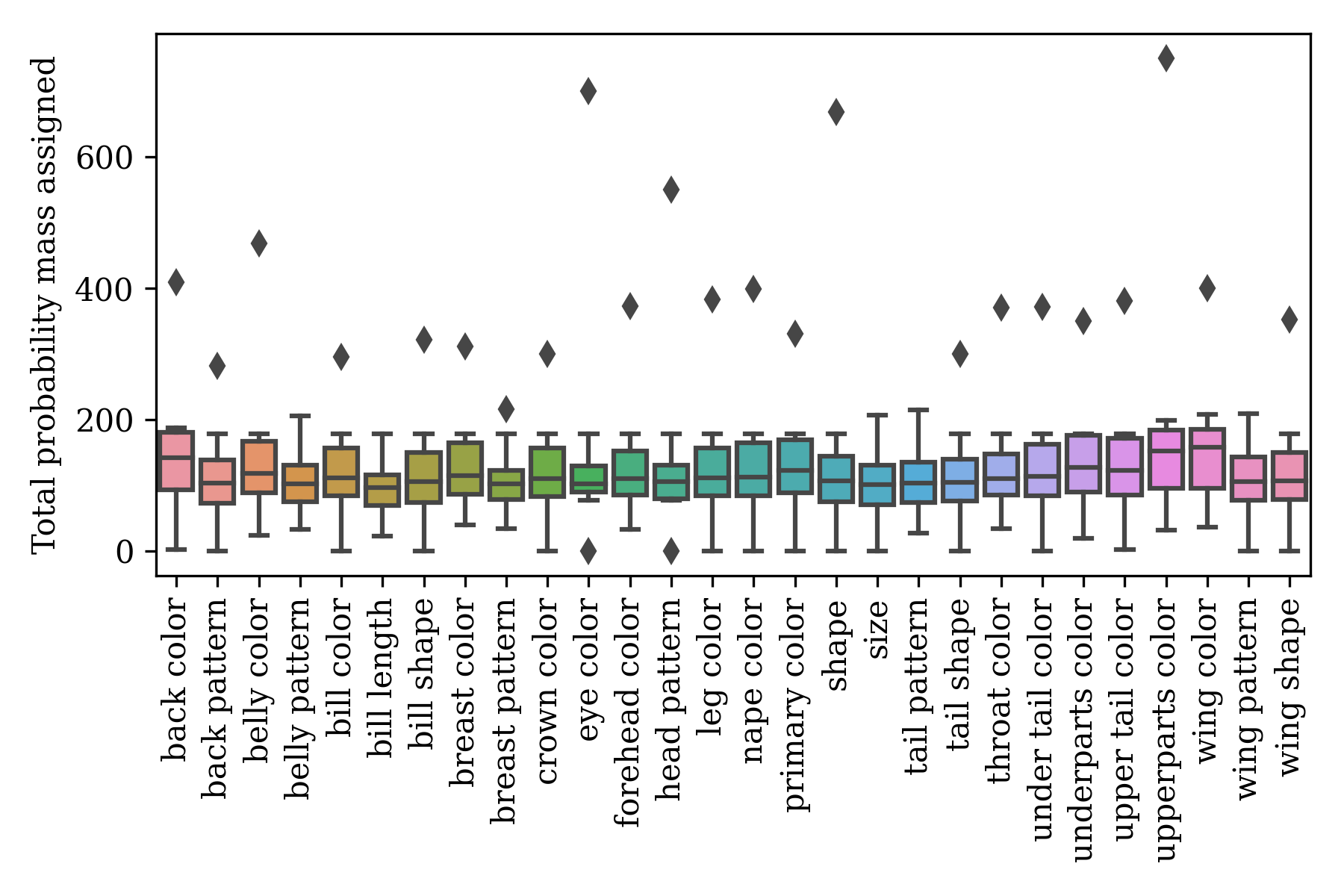}
    \caption{Distribution across images of total probability mass assigned for each concept. There is significant variation in the mean, skew and variance of distributions, showing that different concepts are annotated differently by human annotators.}
    \label{fig:CUBS_prob_mass_concept}
\end{figure}

Further, we investigate the variance in flavor of uncertainty expressed between different concepts. In Figure \ref{fig:CUBS_prob_mass_concept} we plot the distribution of probability mass assigned for each concept. We observe significant variations between concepts, in terms of their mean, variance and skew. Some concepts such as ``eye color'' have a very tight distribution around 100, suggesting those concepts are ``easy'' to annotate. In contrast, some concepts such as ``upperparts color'' show greater variation in the probability mass assigned. These concepts tend to either be color concepts, which can have several correct annotations, or ambiguous concepts like ``size'' which may be harder to annotate correctly.

These observations highlight nuances in the \texttt{CUB-S}\,dataset which aren't present in the original hard annotations. Soft annotations give insights into how humans interpret concepts when labeling and the variation in individual calibration of annotators \citep{selfCiteSoftLabel}. We hope to encourage future work to design ML models and datasets which account for the idiosyncrasies of human uncertain annotations.

\begin{figure}[!htb]
    \centering
    \includegraphics[width=0.9\linewidth]{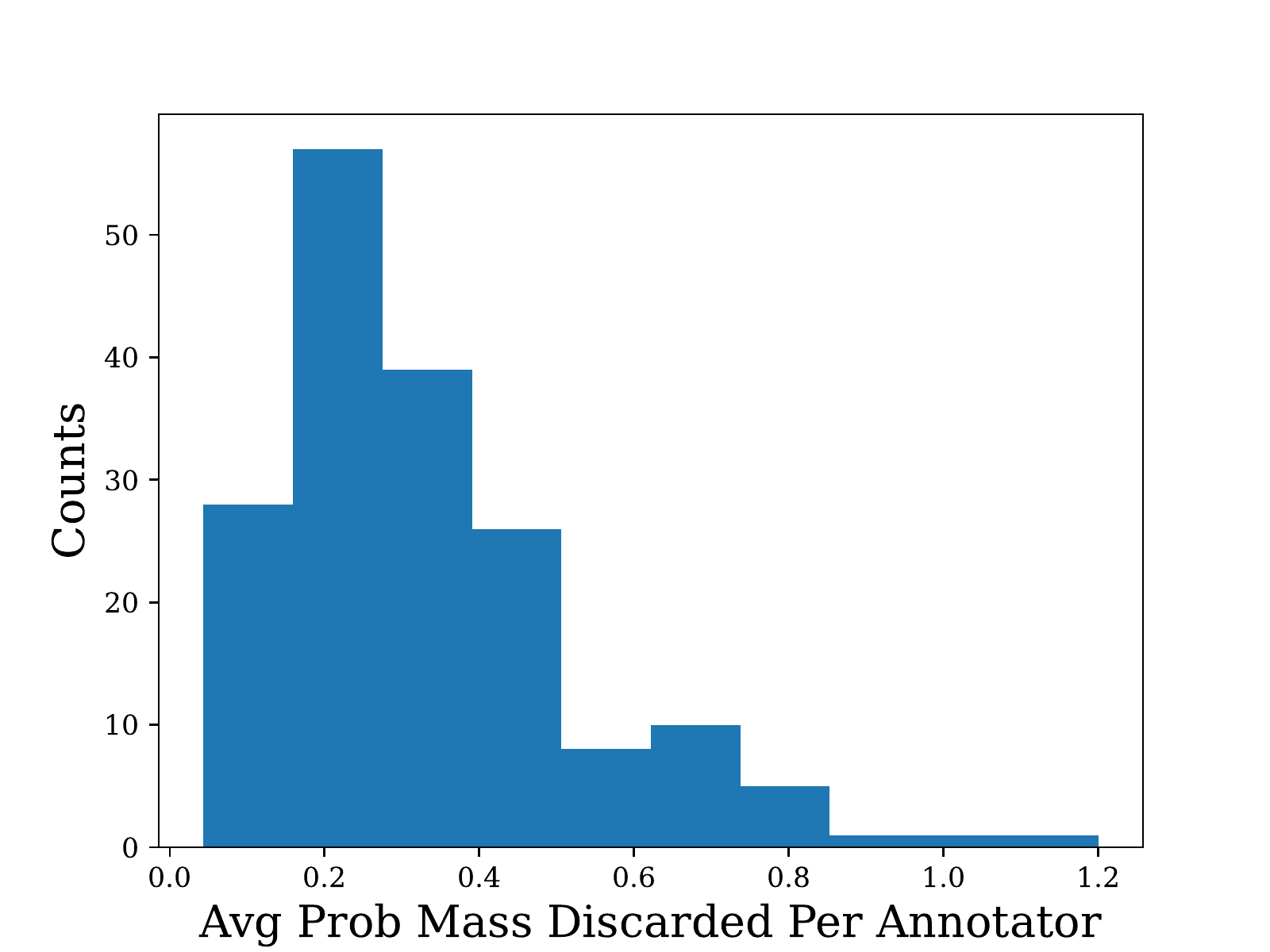}
    \caption{Amount of assigned probability mass discarded per individual when using the popular \citeauthor{koh2020concept} concept filtering (averaged over concept groups). }
    \label{fig:cubs_discarded_mass}
\end{figure}

Additionally, our labels demonstrate potential issues with the concept filtering typically applied on CUB. \citeauthor{koh2020concept} propose a filtering scheme to avoid overly sparse annotations; however, we note that our annotators assign a substantial amount of probability mass to concept attributes which are \textit{filtered out} (see Figure \ref{fig:cubs_discarded_mass}). These data highlight that the filtered out attributes could indeed be missing critical information from people as to what is in the image. 


\begin{figure}[!htb]
    \centering
    \includegraphics[width=0.9\linewidth]{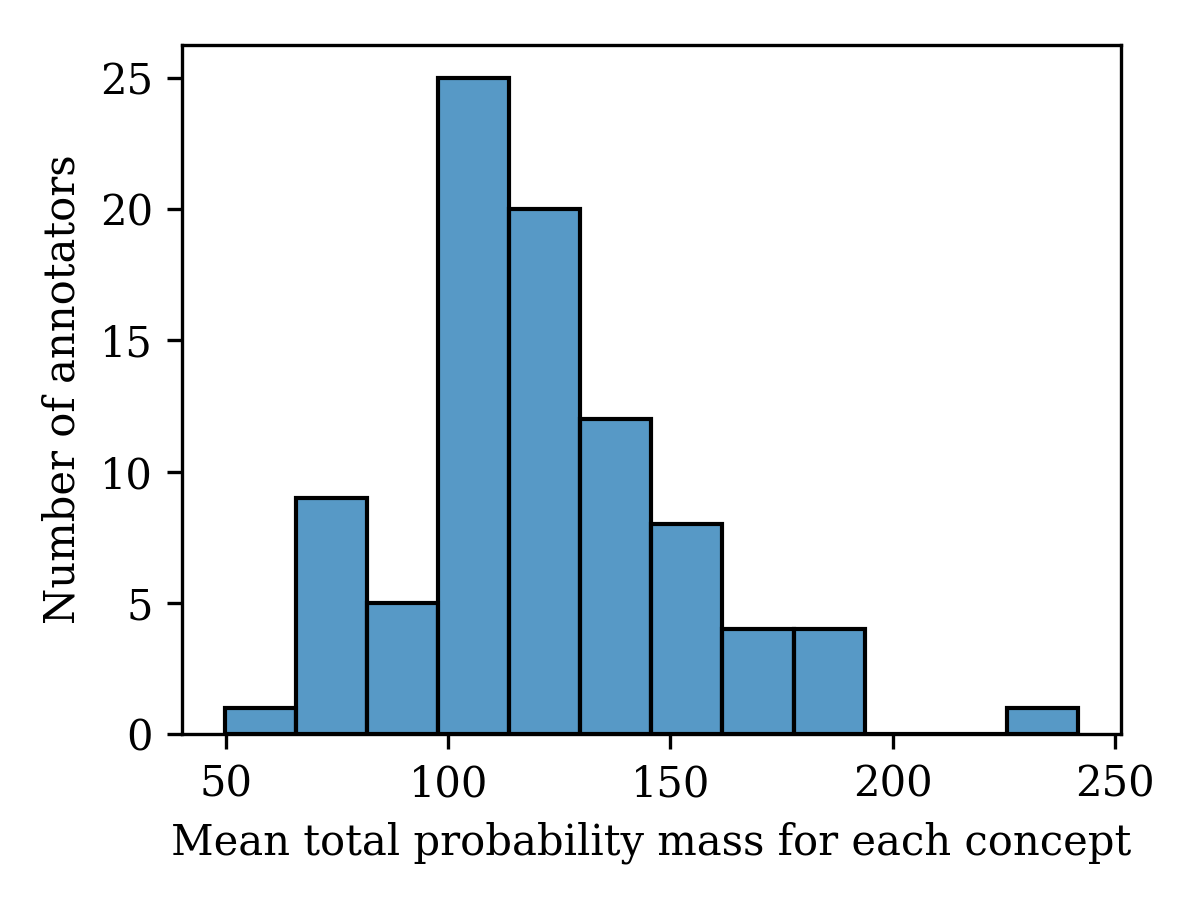}
    \caption{Histogram showing the distribution of mean total probability mass for each concept assigned by each annotator. Most annotators assign approximately 100 probability mass, although there are a significant number which over-assign probability mass.} 
    \label{fig:CUBS_annotator_mass}
\end{figure}

\begin{figure*}
    \centering
    \includegraphics[width=1\linewidth]{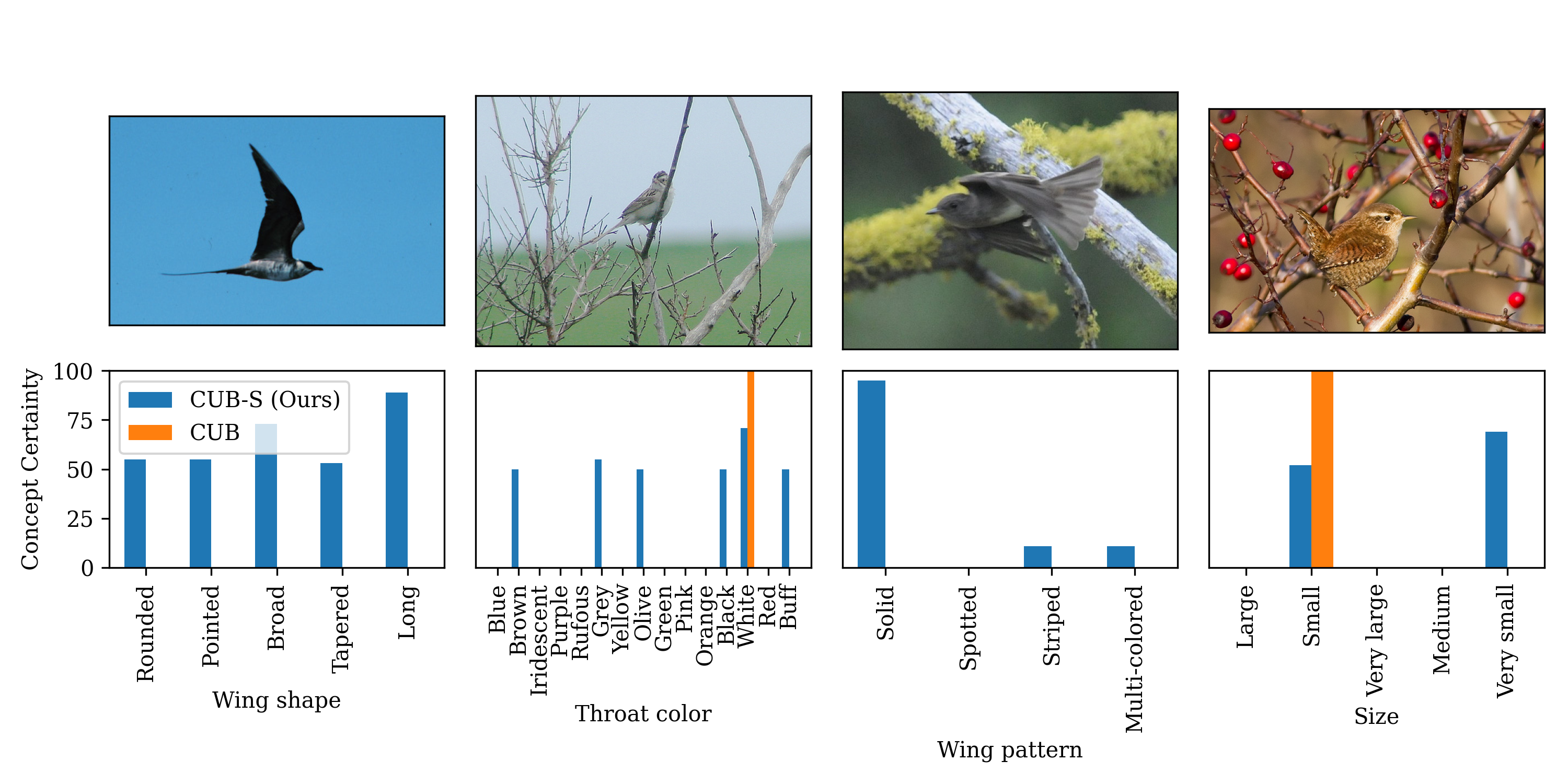}
    \caption{Additional examples showing rich annotations for \texttt{CUB-S} compared to hard assignments in \texttt{CUB}.}
    \label{fig:cubs_examples_2}
\end{figure*}

\begin{figure*}
    \centering
    \includegraphics[width=1\linewidth]{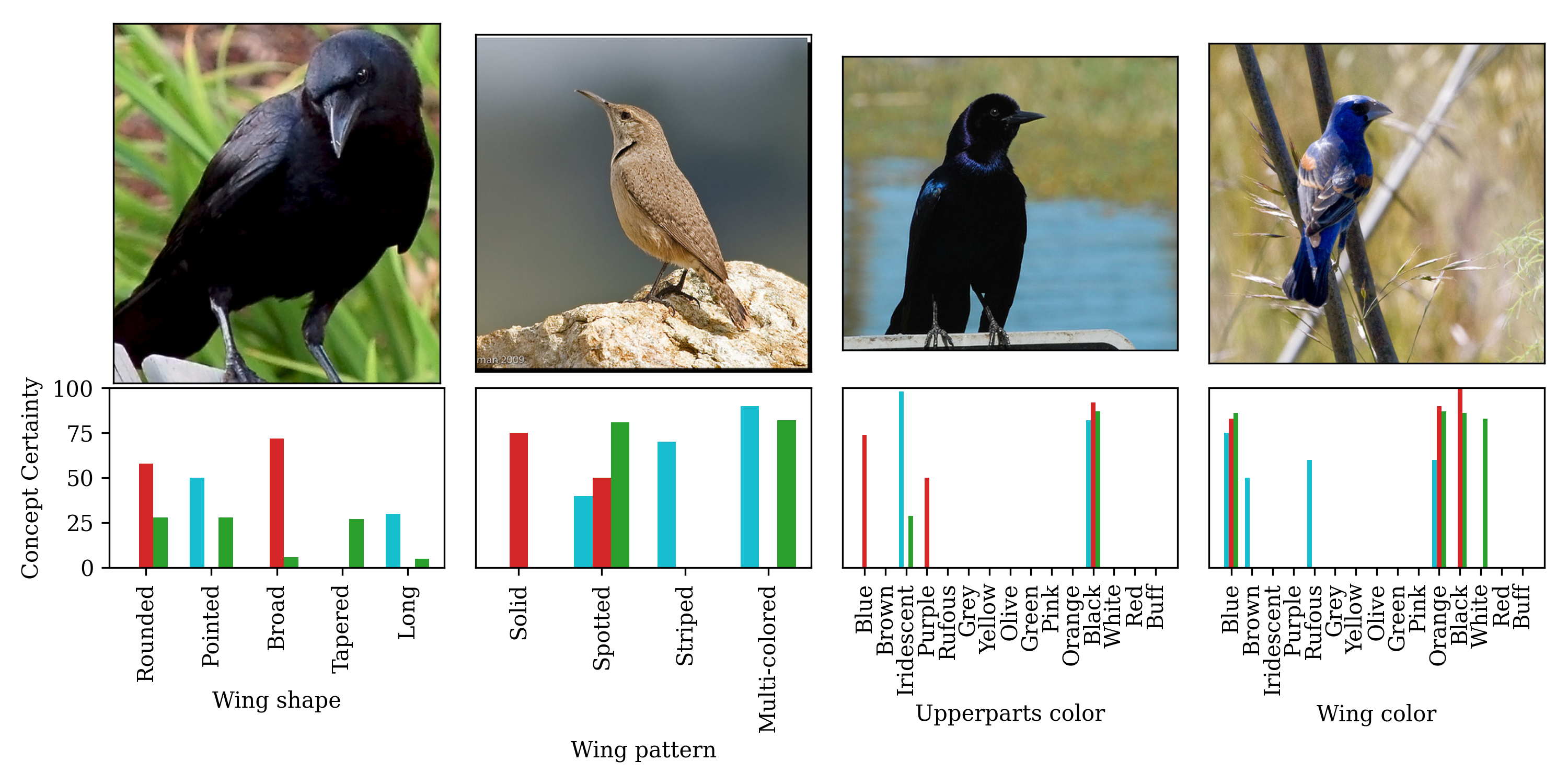}
    \caption{\texttt{CUB-S} Examples where multiple annotators labelled the same image. Each bar color represents a unique annotator for each image. The annotated concepts vary significantly between annotators, especially for challenging concepts such as ``wing shape'' and ``wing pattern''.}
    \label{fig:cubs_examples_annotators}
\end{figure*}

\subsection*{Additional \texttt{CUB} Uncertainty Computational Experiments}

We next include further observations from our computational experiments in \texttt{CUB} and \texttt{CUB-S}.

\subsubsection*{Broad vs. Narrow Uncertainty} 
\begin{figure}[!htb]
    \centering
    \includegraphics[width=0.9\linewidth]{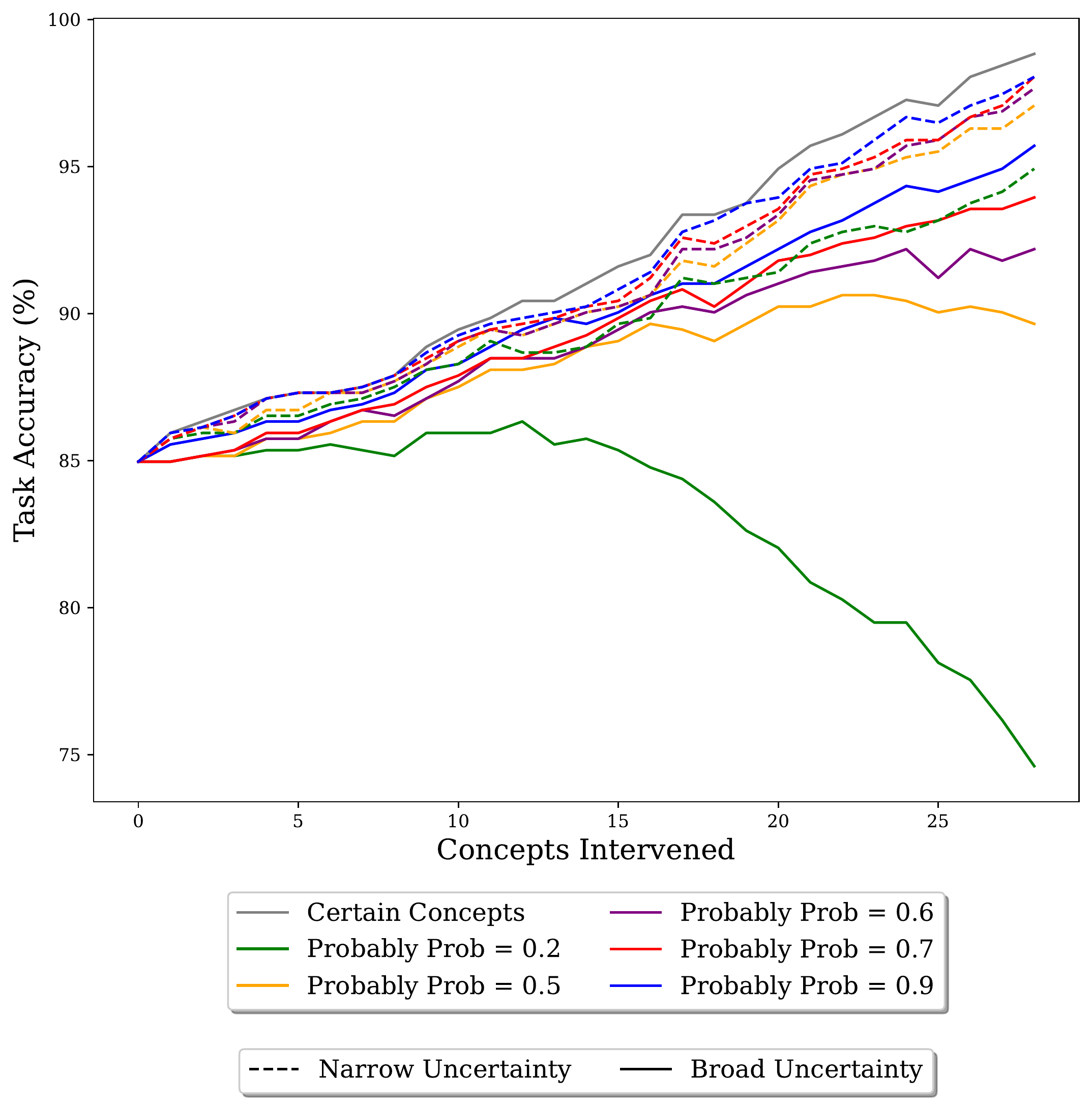}
    \caption{ the efficacy of random interventions on CEMs.}
    \caption{Impact of different ways of distributing the discrete uncertainty over categorical concept groups, selected using Random intervention policies on CEMs. }
    \label{fig:random_smooth_off}
\end{figure}

We demonstrate the sensitivity of concept-based systems to broad versus narrow uncertainty under the Random policy (see Figure \ref{fig:random_smooth_off}), further highlighting that the method of distributing uncertainty through discrete confidence scores matters impacts intervention efficacy.

\subsubsection*{Individual- vs Population-Level Uncertainty} 
\begin{figure}[!htb]
    \centering
    \includegraphics[width=0.9\linewidth]{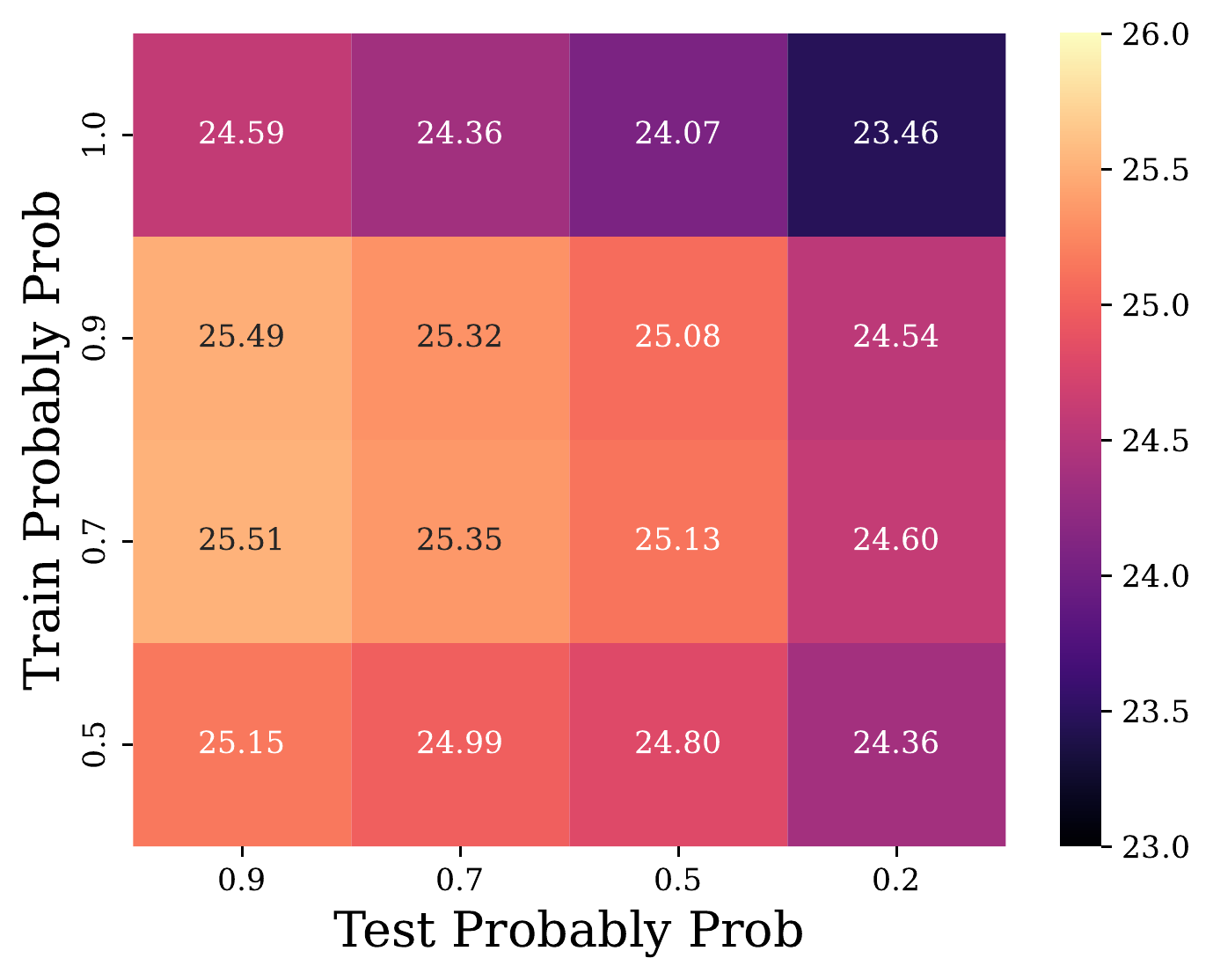}
    \caption{Training with a moderate level of (aggregate/population-level) uncertainty improves robustness under test-time uncertainty; as measured by AUC between intervention-accuracy curve. Higher is better.} 
    \label{fig:cub_train_uncertainty}
\end{figure}

As noted, whether or not we intervene with individual or population-level annotations matters (see Section 5.2.3), and we see in Figure \ref{fig:cub_train_uncertainty} that training and then intervening with population-level annotations yields the best performance. These observations are relevant not only to ML practioners who work with \texttt{CUB}, but broadly in annotation-design and questions around who and how many annotators should we elicit from.

\begin{figure}[!htb]
    \centering
    \includegraphics[width=0.9\linewidth]{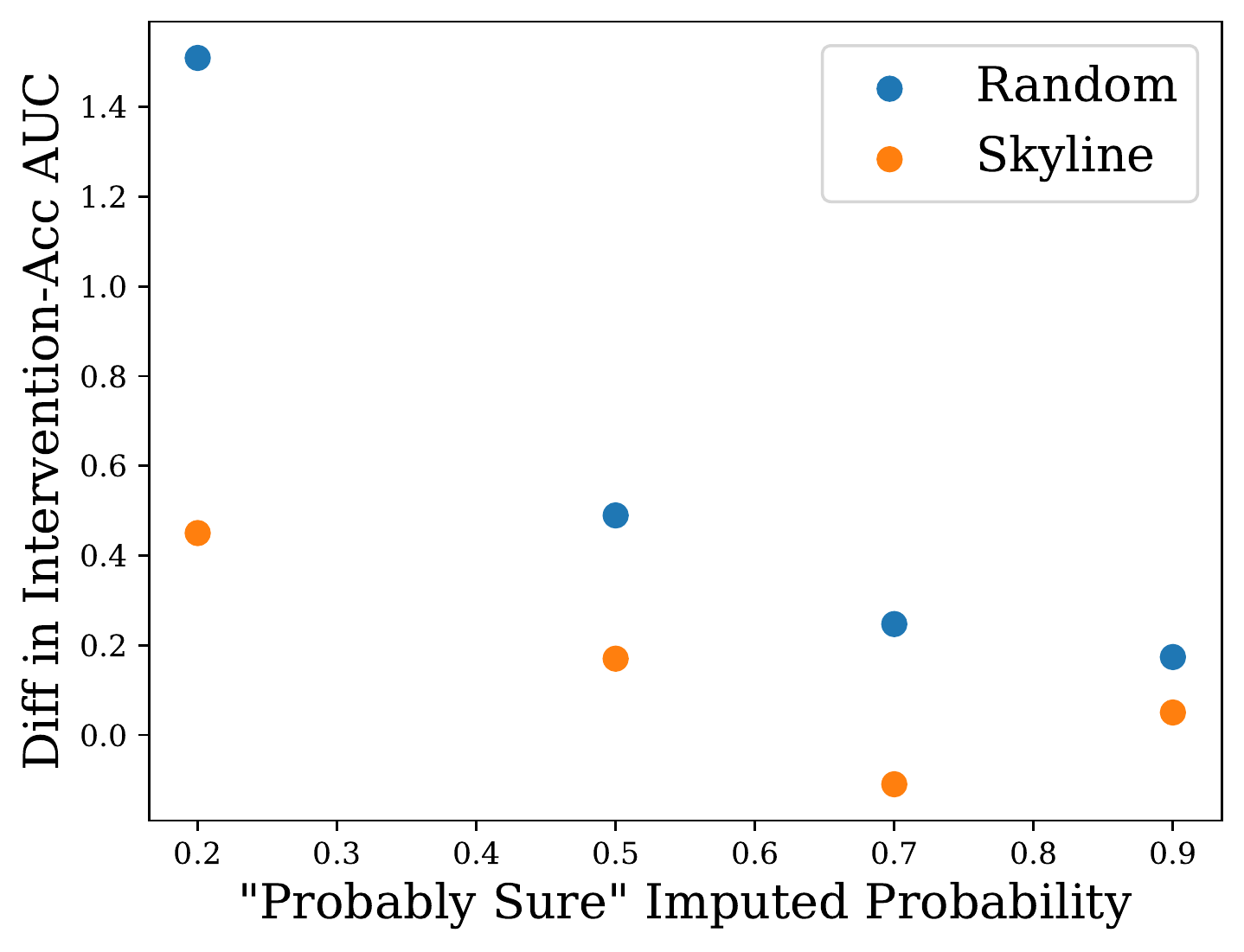}
    \caption{It matters whether or not we use \textit{instance-level, individual} annotator uncertainty, or average over many individuals' uncertainty. Averaging improves the stability of interventions; but in practice, we may only have a single individual who can provide their uncertainty. We find sizeable differences in the intervention efficacy when using averaged uncertainty for both Skyline and Random.}
    \label{fig:avgd_unc}
\end{figure}

\subsubsection*{CBMs and Simulated Uncertainty}
\begin{figure}[!htb]
    \centering
    \includegraphics[width=0.9\linewidth]{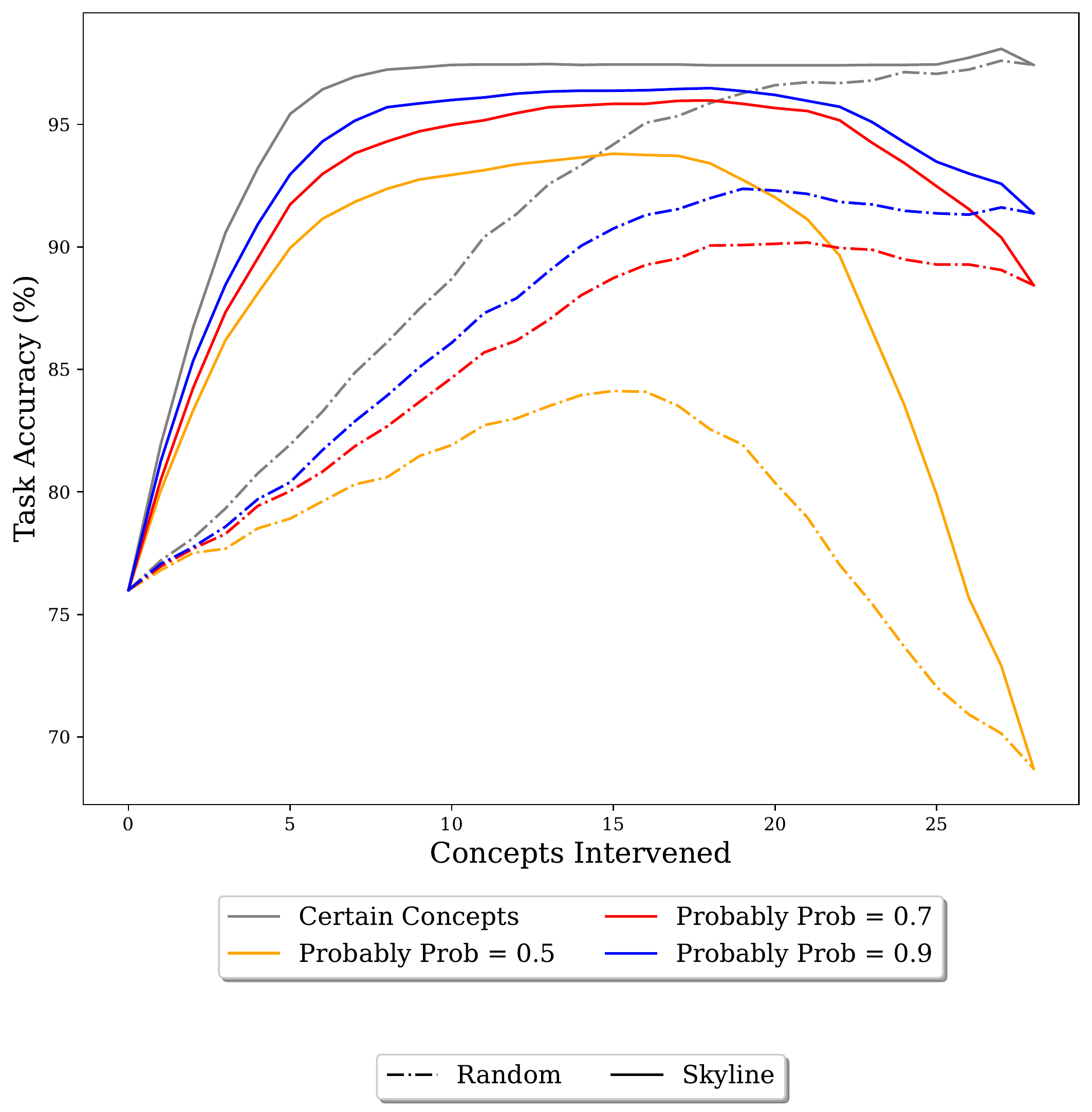}
    \caption{CBMs struggle to handle uncertainty in CUB as well and are comparatively worse than CEMs.}
    \label{fig:cbm_cub_unc}
\end{figure}

Further, we concretize why we focus on CEMs in the bulk of this work. CBMs severely struggle under test-time uncertainty when dealing with concept-incomplete datasets (see the \texttt{UMNIST} section of this Supplement) and in-the-wild uncertainty (see Figure \ref{fig:cbm_cub_unc}). 

\subsubsection*{Skyline Selections Reveal ``Helpful'' and ``Harmful'' \texttt{CUB-S} Annotations} 

\begin{figure*}[!htb]
    \centering
    \includegraphics[width=0.9\linewidth]{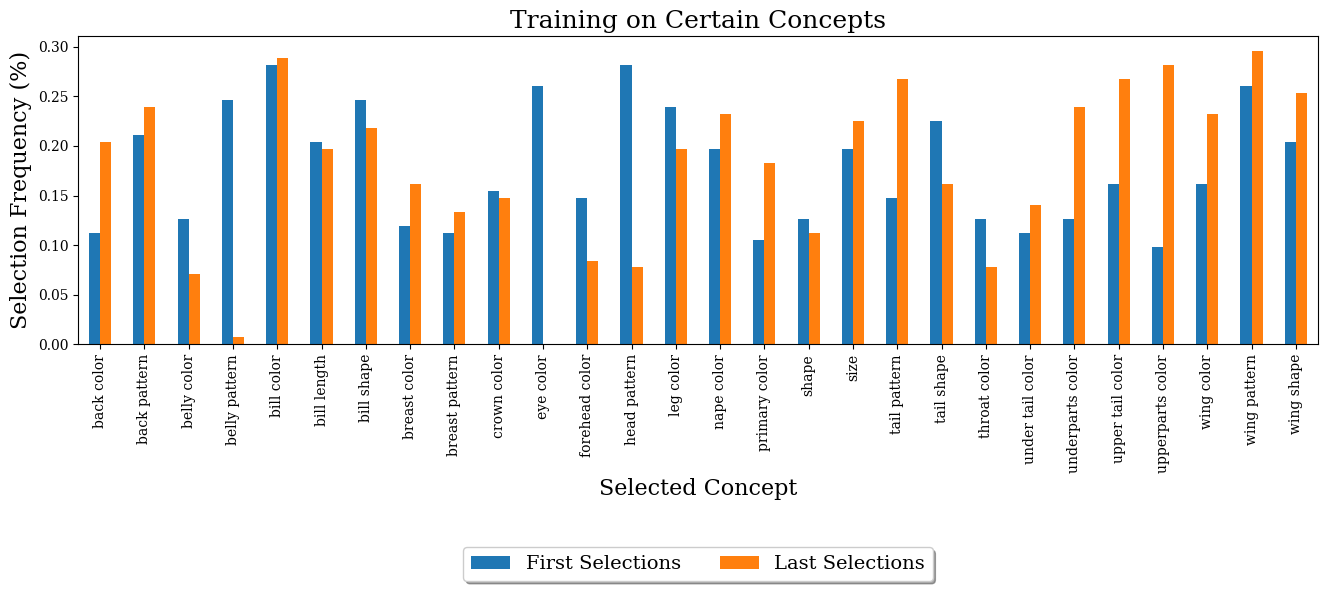}
    \caption{Skyline selections for CEM run on \texttt{CUB-S} reveal when human uncertainty elicitation is helpful (versus harmful). Proportion of selections for each concept being in the first or last 5 selections by Skyline. CEM trained on certain concepts.}
    \label{fig:skyline_cem_sels_none}
\end{figure*}

\begin{figure*}[!htb]
    \centering
    \includegraphics[width=0.9\linewidth]{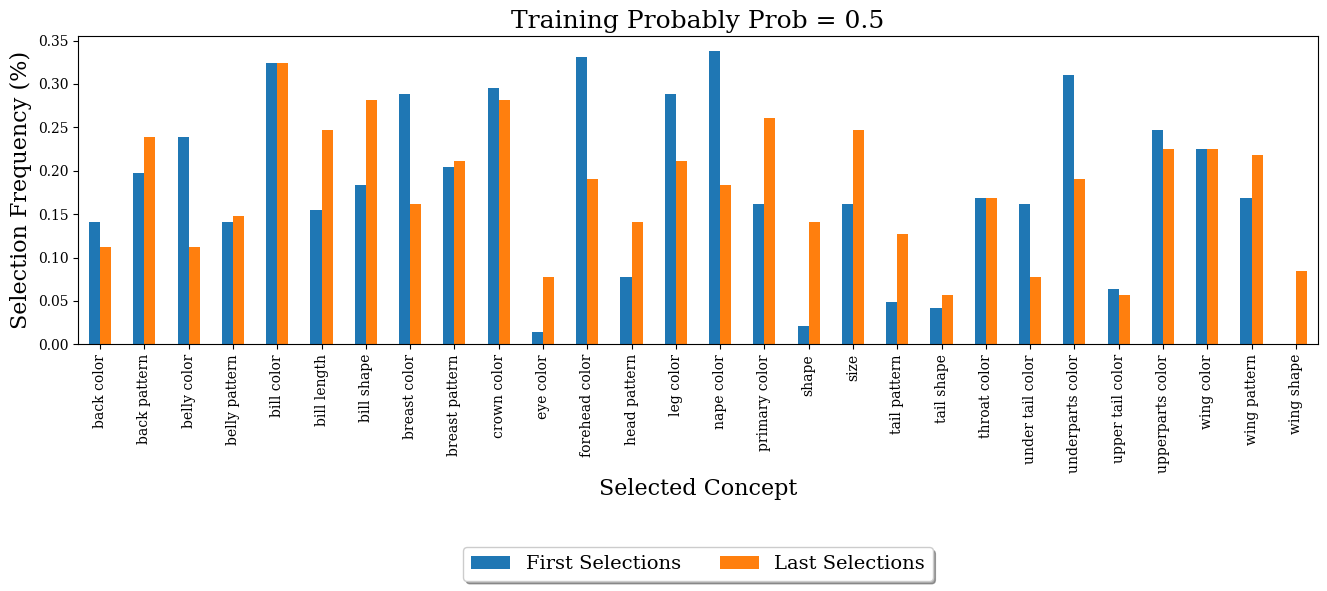}
    \caption{Skyline selections for CEM trained on uncertain concepts (where the imputed ``Probably'' probability is set to 0.5). Population-level broad uncertainty labels used.}
    \label{fig:skyline_cem_sels0.5}
\end{figure*}

\begin{figure*}[!htb]
    \centering
    \includegraphics[width=0.9\linewidth]{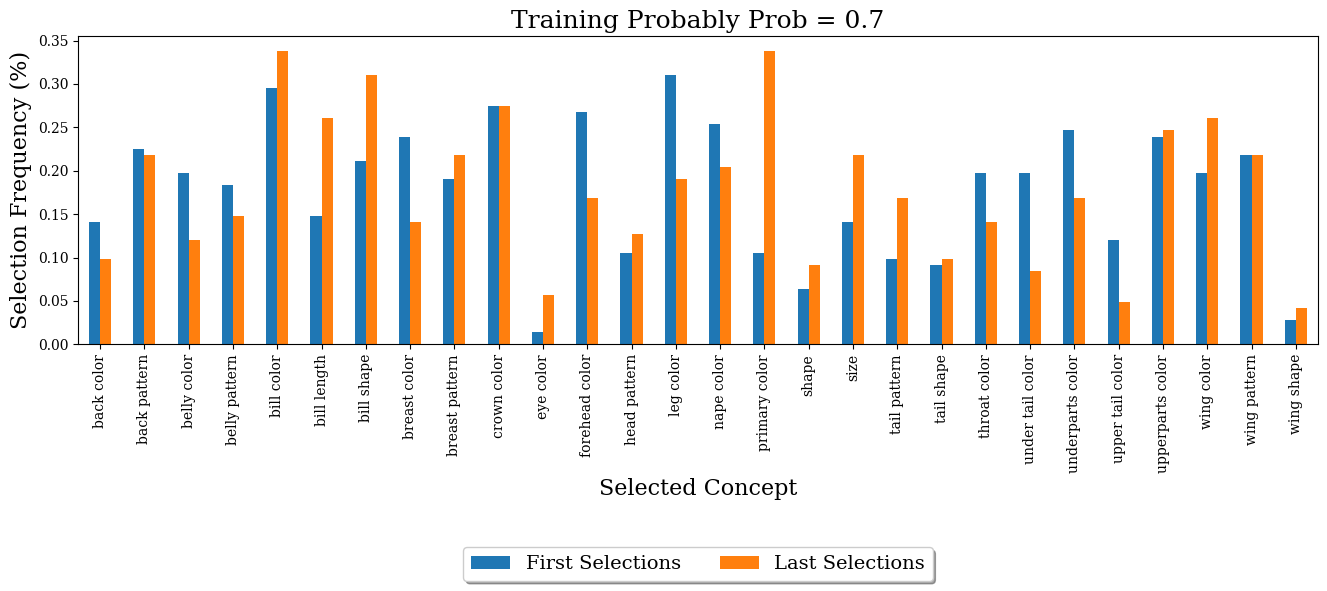}
    \caption{Skyline selections for CEM trained as in Figure \ref{fig:skyline_cem_sels0.5}, but with the imputed ``Probably'' probability set to 0.7.}
    \label{fig:skyline_cem_sels0.7}
\end{figure*}

\begin{figure*}[!htb]
    \centering
    \includegraphics[width=0.9\linewidth]{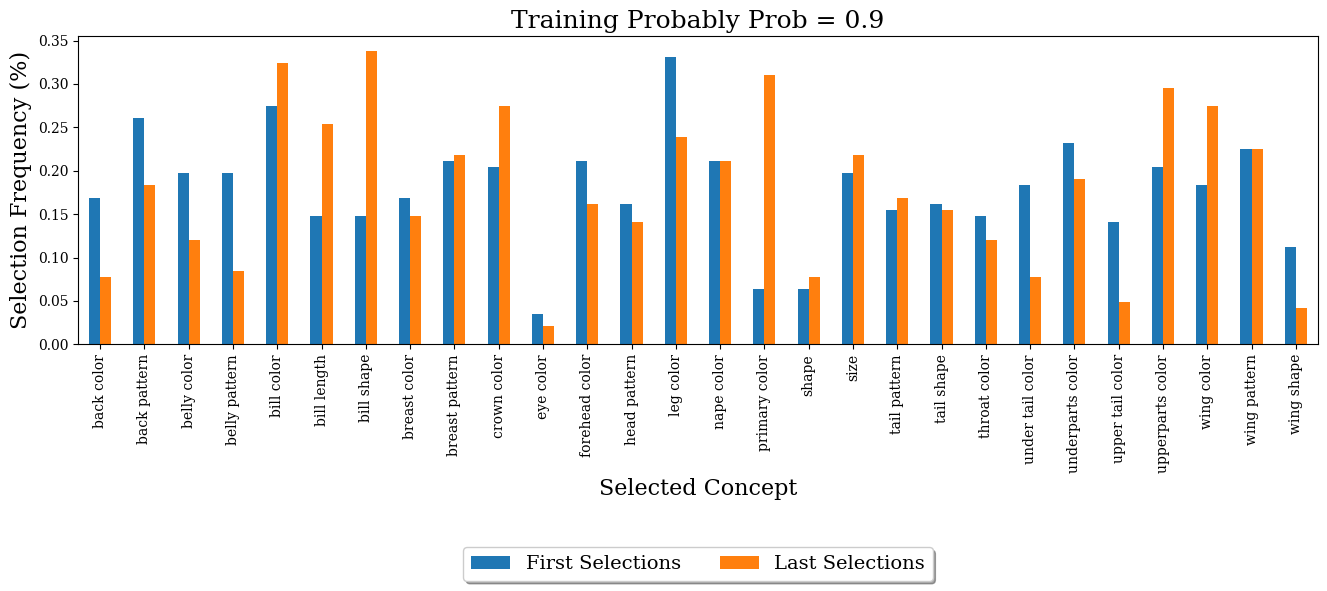}
    \caption{Skyline selections for CEM trained as in Figure \ref{fig:skyline_cem_sels0.5}, but with imputed ``Probably'' probability set to 0.9.}
    \label{fig:skyline_cem_sels0.9}
\end{figure*}

 As seen in Figure \ref{fig:cubs_policies}, Skyline rapidly improves by selecting ``good'' uncertain annotations; however, the final selections hamper performance. We depict the proportion of selections for each concept being in the first or last 5 selections by Skyline. Avoiding selecting the examples in the last 5, e.g., ``upperparts'' color, offer promising directions for future policy design and investigation into when and why humans are good uncertain annotators. Interestingly, we observe differences in which concepts are preferred depending on whether the model was trained without (Figure \ref{fig:skyline_cem_sels_none}) or with uncertainty in the concepts at training time (i.e., Figures \ref{fig:skyline_cem_sels0.5}, \ref{fig:skyline_cem_sels0.7}, \ref{fig:skyline_cem_sels0.9}). 
 
\end{document}